\documentclass[lettersize,journal]{IEEEtran}
\usepackage{algorithmic}
\usepackage{array}
\usepackage[caption=false,font=normalsize,labelfont=sf,textfont=sf]{subfig}
\usepackage{textcomp}
\usepackage{stfloats}
\usepackage{url}
\usepackage{verbatim}
\usepackage{graphicx}
\usepackage{cite}
\usepackage{amsmath,amssymb,amsfonts}
\usepackage[ruled,vlined,linesnumbered]{algorithm2e}
\usepackage{comment}
\usepackage{multirow}
\usepackage{tabularx}
\usepackage{makecell}
\usepackage[table,xcdraw]{xcolor}
\usepackage{enumitem}
\usepackage{threeparttable}
\usepackage[disable]{todonotes}
\usepackage{booktabs}
\newcolumntype{Y}{>{\centering\arraybackslash}X}
\usepackage{xpatch}
\makeatletter
\xpatchcmd{\IEEEbiography}
  {\vskip \@IEEEBIOskipN plus 1fil minus 0\baselineskip}
  {\vskip \@IEEEBIOskipN plus 0pt minus 0\baselineskip}
  {}{\typeout{IEEEbiography patch failed}}
\makeatother
\def\BibTeX{{\rm B\kern-.05em{\sc i\kern-.025em b}\kern-.08em
    T\kern-.1667em\lower.7ex\hbox{E}\kern-.125emX}}

\newif\ifshowrevisioncomments
\showrevisioncommentsfalse
\ifshowrevisioncomments
\newcommand{\revision}[1]{\textcolor{blue}{{#1}}}
\newcommand{\deletion}[1]{\textcolor{red}{{#1}}}
\else
\newcommand{\revision}[1]{#1}
\newcommand{\deletion}[1]{}
\fi

\begin{document}
\title{Spotting Setting-Related UI Display Bugs in Android Apps}
\author{Huaxun Huang, Wu Liu, Jiahao Gu, Rongxin Wu
\thanks{Manuscript created January, 2025; Huaxun Huang, Wu Liu, Jiahao Gu and Rongxin Wu are with School of Informatics (National Characteristic Demonstration Software School), Xiamen University, Xiamen, 361005, China (e-mail: huanghuaxun@xmu.edu.cn; liuwu@stu.xmu.edu.cn; gujiahao@stu.xmu.edu.cn; wurongxin@xmu.edu.cn). Huaxun Huang and Rongxin Wu are also with Fujian Engineering Research Center of High-Performance Intelligent Computing Systems. Rongxin Wu is the corresponding author.

}}

\markboth{IEEE Transactions on Software Engineering}%
{Huang et al.: Spotting Setting-Related UI Display Bugs in Android Apps}

\maketitle

\begin{abstract}
Android provides a wide range of system settings that allow users to control the runtime behaviors of apps, such as screen rotation and UI display. However, setting-related bugs occur when developers do not fully align their apps with the extensive range of system settings that users can define. These bugs can commonly affect apps' UI, causing setting-related UI display (SUD) bugs that negatively impact user experience. While existing research has explored automated detection of SUD bugs, these approaches often suffer from false negatives. This limitation stems from an incomplete understanding of how app components should adapt UI elements to diverse system settings. To address this gap, we conducted an empirical study to identify common patterns of unexpected setting adaptations that result in SUD bugs. These patterns then served as the test oracle for our proposed automated tool, SUDFinder. To ensure the test coverage, SUDFinder injects a test activity to visually render the XML configuration files of each UI page. We evaluated SUDFinder on 29 popular, open-source apps on F-Droid and found that it effectively identifies 98 previously unknown SUD bugs, achieving a precision of 0.76. So far, 67 have been confirmed and 37 have been fixed by the app developers.
\end{abstract}


\section{Introduction}
\IEEEPARstart{A}{ndroid}, a popular mobile OS, offers various predefined components that developers utilize through XML to build apps' UI. A study of 200 top-ranked apps~\cite{huang2021characterizing} found an average of 25,991.6 XML elements across 663.1 layout files per app, highlighting the complexity and extensive use of XML in defining mobile UIs.

The Android system provides a range of user-configurable system settings to control the runtime behavior of apps. For example, users can change the system language, adjust the system theme, grant or revoke app permissions, or adjust screen orientation. However, setting-related bugs occur when app developers fail to adapt their apps to different types of system settings. Such setting-related bugs often lead to both app crashes and non-crashing functional bugs. We refer to these as \textbf{S}etting-Related \textbf{U}I \textbf{D}isplay (SUD) bugs. As revealed by Sun et al.~\cite{sun2021understanding, sun2023characterizing}, SUD bugs are common, accounting for 20.3\% (218/1074) of the real-world setting-related bug dataset.

\begin{figure}[t]
    \centering    \includegraphics[width=0.48\textwidth]{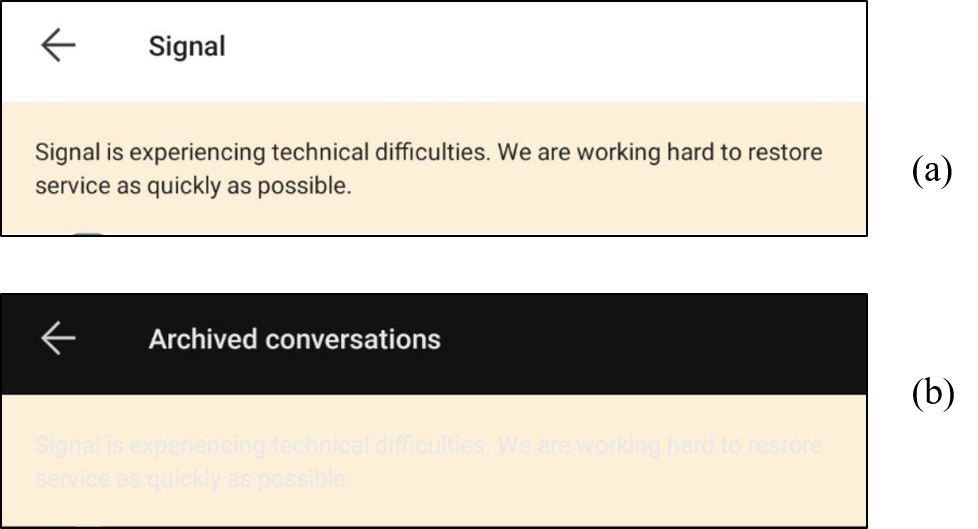}
    \caption{A real-world SUD bug is reported in Signal-Android issue \#11258~\cite{signal11258}, where the app developers set the text color in dark mode with insufficient contrast against the background, making the text difficult to read.}
    \label{fig:motivating_example}
\end{figure}

\begin{figure}[t]
        \includegraphics[width=0.55
        \textwidth]{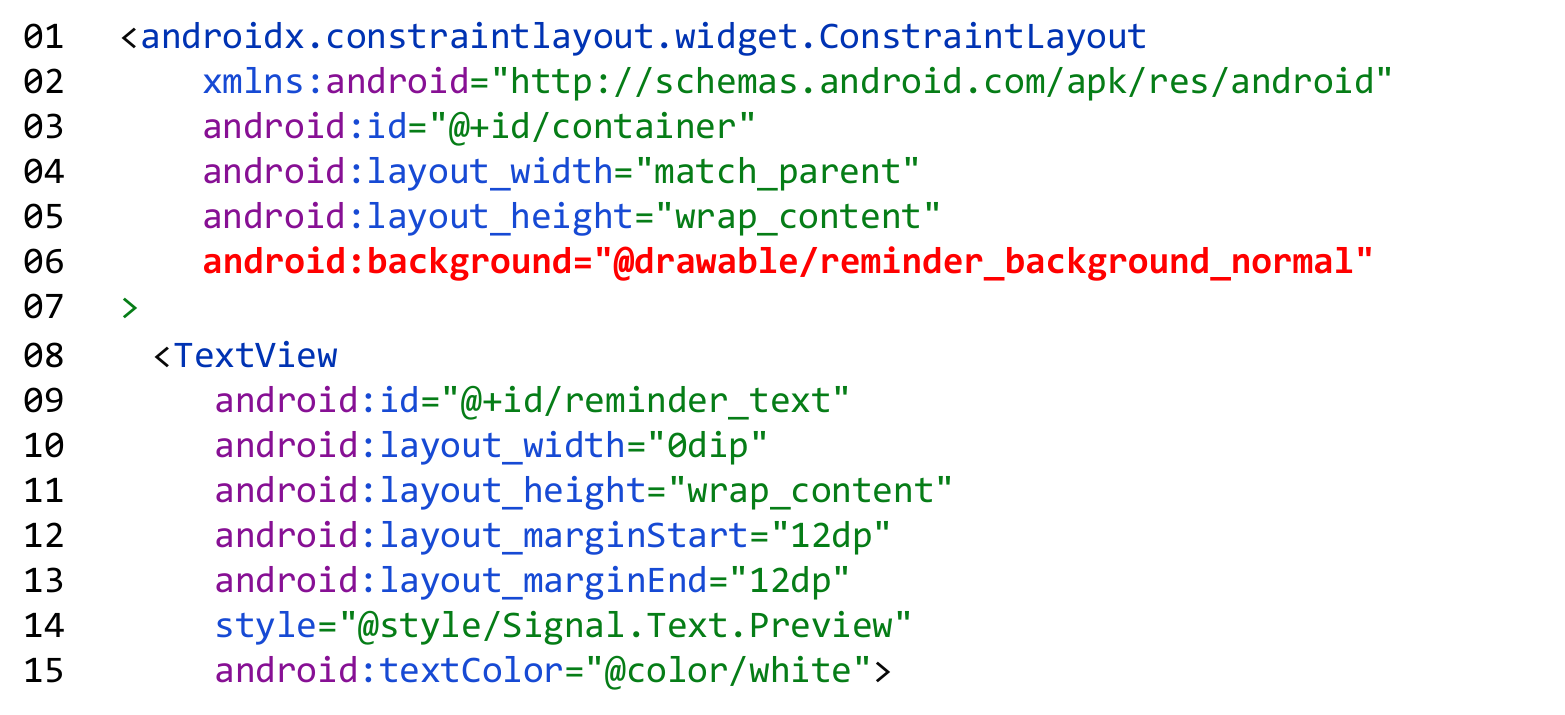}
        \caption{The root cause of the SUD bug reported in Signal-Android \#11258~\cite{signal11258}.}
        \label{fig:patch}
\end{figure}

Figure \ref{fig:motivating_example} illustrates a real-world SUD bug that impacted the app’s functionality. The bug is originally reported in Signal-Android issue \#11258~\cite{signal11258}, a free, open-source private messenger with over 100 million installations on Google Play. In Figure \ref{fig:motivating_example}(a), the FrameLayout is displayed under the default configuration, where the banner text is set to black and the background is antique white. However, when the user changes the system setting to dark mode, it can be seen that the developers did not update the color of the background, but modified the text from black to white. As a result, this causes an SUD bug in which the banner text becomes invisible to users in dark mode (see Figure \ref{fig:motivating_example}(b)). Figure \ref{fig:patch} shows the root cause of the bug illustrated in Figure \ref{fig:motivating_example}. As shown on Line 7, the developers hard-coded the same background color for both light and dark modes, resulting in insufficient color contrast when the system changes the text color in the dark mode.


From the perspective of app developers, detecting SUD bugs is a challenging task due to the large search space consisting of UI components~\cite{huang2021characterizing} and system settings~\cite{sun2021understanding, sun2023characterizing}. Although Google has provided some official guidelines for developers on adapting the apps' UI to different system settings \cite{guide}, developers often neglect the content described in these official guidelines. Even popular, well-tested apps may be unexpectedly affected due to inadequate consideration of various setting changes~\cite{sun2021understanding,sun2023characterizing}. Manually testing these bugs requires crafting and maintaining the test scripts, which are subsequently executed automatically to explore the apps' UI and capture corresponding screenshots across system settings, followed by a manual comparison of each component to check for potential bugs. This manual process is both time-consuming and labor-intensive. Consequently, there is a pressing need for an automated approach to detect SUD bugs and alleviate the burden of manual testing~\cite{linares2017howdodevtest,mahmud2024apptestingpractices}.

However, developing an automated method for the detection of SUD bugs faces the following two key challenges: covering all potential UI components and establishing a robust test oracle. The first ensures broad bug detection, while the second confirms if a UI component has SUD bugs. Each system setting requires a tailored oracle, considering the app's specific context for accurate validation.

Numerous approaches~\cite{sun2023characterizing,guo2022detecting, zhou2023ddldroid, sun2021understanding,lint,su2022metamorphosis,alshayban2022accessitext,  sadeghi2017patdroid,riganelli2020data,eler2018automated, salehnamadi2021latte,salehnamadi2022groundhog,accessibilityscanner,lu2019preference, gu2025characterizing, xiao2026ldmdroid} have been proposed to automatically detect setting-related bugs in Android apps.
Existing static-based approaches \cite{sun2023characterizing, guo2022detecting, zhou2023ddldroid, lint} primarily work at the API invocation level, checking whether an API is called under incorrect system settings or if an API sequence conforms to the specifications of those settings. However, the above approaches do not target SUD bugs, which are primarily located in XML files and involve the hierarchical structure of XML elements and attribute configurations (See Section \ref{sec:empirical}).  
The test case generation process of existing dynamic-based approaches~\cite{accessibilityscanner, sun2021understanding, su2022metamorphosis, alshayban2022accessitext, sadeghi2017patdroid,riganelli2020data, lu2019preference} relies on random testing, which is inefficient and can require prohibitively long execution times to achieve high coverage of all UI components. Moreover, the test oracles are incomplete in covering the expected adaptations of system settings, resulting in false negatives. For example, the bug illustrated in Figure \ref{fig:motivating_example} involves unexpected text color when changing the system to the dark mode, which has not been addressed by existing work (See Section \ref{sec:motivation}).

To bridge the aforementioned research gap, we conducted an empirical study based on real-world SUD bugs. Our study identified common patterns of unexpected UI adaptations that lead to SUD bugs. Based on these findings, we further propose SUDFinder, an automated approach for detecting SUD bugs in Android apps. The test oracle of SUDFinder integrates the common patterns identified in our empirical study, which indicate that \textit{UI components with similar visual features tend to adapt in similar ways to setting changes} (e.g., inconsistencies in text color changes and background changes, as shown in Figure~\ref{fig:motivating_example}).
To cover more UI components in the app under test, SUDFinder performs bug detection through the app's UI XML configuration files, thus avoiding the need to dynamically explore the app's UI through building test cases to achieve high UI component coverage. Although this process ignores the app code that might dynamically change the app's UI, we have empirically found that \textit{SUD bugs are commonly incurred because app developers fail to adapt the XML configuration files of UI components across different system settings} (see Section \ref{sec:empirical}). 

We implemented SUDFinder and conducted evaluations on 29 open-source popular Android apps in F-Droid~\cite{fdroid}. The results show that SUDFinder successfully detected 98 bugs with a precision of 0.76. SUDFinder uncovers 51 bugs that cannot be detected by state-of-the-art approaches~\cite{sun2023characterizing,escobar2020empirical, chatgpt}. We also submitted the bug reports of previously-unknown SUD bugs to the original app developers. Encouragingly, 67 bugs have been confirmed and 37 bugs have been merged by the app developers, demonstrating the usefulness of SUDFinder.

In summary, this paper makes the following contributions:
\begin{itemize}[leftmargin=3ex]
    \item To the best of our knowledge, we conducted the first empirical study on SUD bugs to understand the common code patterns and bug manifestation.
    \item We proposed SUDFinder, an approach that relies on both XML code and dynamic information to detect SUD bugs in Android apps.
    \item We systematically evaluated the effectiveness and usefulness of SUDFinder. The results show that SUDFinder can effectively help detect previously-unknown SUD bugs, while receiving positive feedback from app developers.
\end{itemize}

\section{Background \& Motivations}

\subsection{UI Rendering in Android Apps}

Android provides a powerful and flexible framework that allows developers to build visually appealing and highly interactive UI. The UI rendered by an Android app is composed of a tree structure of views, which are the basic building blocks of the Android UI. Initially, developers use XML files to define the tree structure of the UI. These XML files are stored in the \texttt{res/layout} directory and are transformed into actual UI components at runtime. The attributes of the views (such as view ID, layout height, and text content) defined in XML configuration files determine their initial appearance and behavior, such as the position and size of buttons, text views, and image views. Furthermore, developers can invoke \texttt{findViewById} in the app code to obtain these views and adjust the visual appearance of views dynamically.

\subsection{Multi-Modal Large Language Model}
Multi-modal large language models (MLLMs) are advancing rapidly,  opening new frontiers for mobile application development and automation. Recent models, such as OpenAI’s GPT-4o and Alibaba's Qwen-VL, demonstrate strong proficiency in interpreting images and natural language instructions to generate corresponding textual and visual outputs. This progress suggests that MLLMs are transitioning from simple image recognition to more complex visual  reasoning: they are increasingly adept at semantic interpretation and  reasoning about visual inputs, effectively ``understanding'' depicted  content and connecting it to user intent more intelligently.

\begin{figure}
    \centering
    \includegraphics[width=0.97\linewidth]{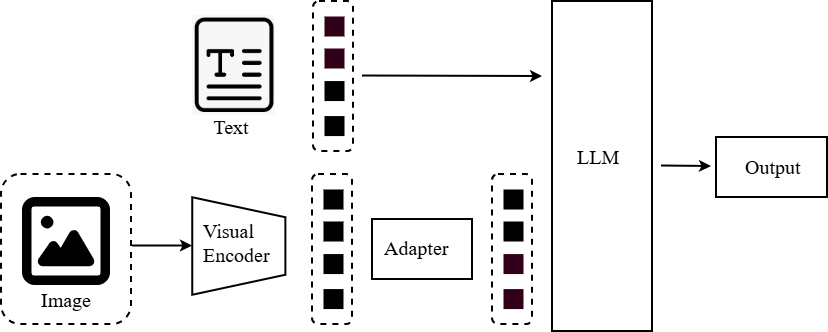}
    \caption{General architecture of MLLMs}
    \label{fig:mllm_overview}
\end{figure}

As illustrated in Figure \ref{fig:mllm_overview}, a general MLLM architecture typically comprises three primary components: a  visual encoder, a projection adapter, and the core MLLM itself~\cite{yin2024survey}. The visual encoder first processes the input image, converting it into a sequential representation of visual tokens. These tokens are subsequently fed through the projection adapter, which maps them into the same embedding space as the model's textual inputs. This integration creates a unified representation for images and text, allowing the core MLLM to process and reason jointly over visual and textual information to generate coherent responses using its inherent language modeling capabilities.

Prior works have used MLLMs as effective tools for UI analysis, applying them to tasks such as UI testing \cite{yu2025visionbasedgui} and UI bug detection \cite{liu2025seeingisbelieve, ju2024studyofusingmllm}. For instance, researchers have leveraged MLLMs to interpret screenshot  images and detect both UI display and UI logic bugs based on visual cues \cite{ju2024studyofusingmllm}. These works highlight the MLLMs' multimodal fusion capability for detecting UI issues such as layout confusion, color discordance, or unclear fonts  based on the screenshot \cite{yu2025visionbasedgui}, which motivates us to employ MLLMs as a baseline for identifying potential SUD bugs.

\begin{figure}[t]
    \centering
    \includegraphics[width=0.48\textwidth]{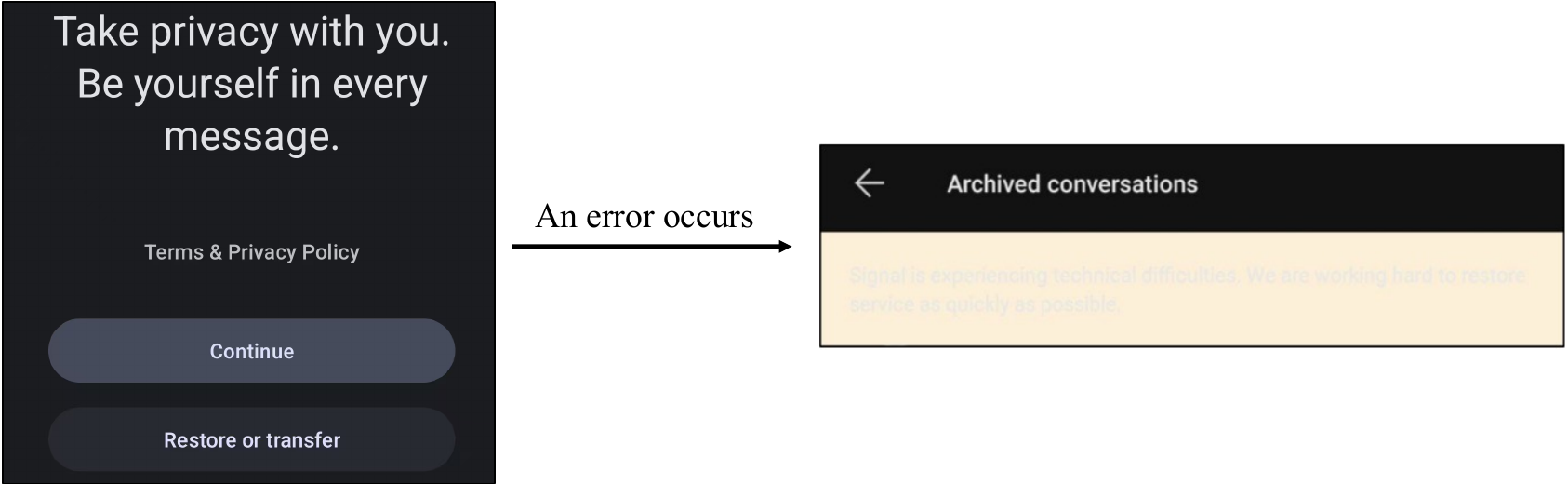}
    \caption{The reproduction steps for Signal-Android issue \#11258~\cite{signal11258}}
    \label{fig:reproduction_step}
\end{figure}

\subsection{Motivating Example}
\label{sec:motivation}

In this section, we use the bug illustrated in Figure \ref{fig:motivating_example} to highlight the limitations of existing approaches and motivate the design of our approach.
\subsubsection{Setup}

We employed SetDroid~\cite{sun2021understanding,sun2023characterizing} and dVermin~\cite{su2022metamorphosis} to analyze the UI depicted in Figure \ref{fig:motivating_example}. SetDroid identifies bugs by comparing app behavior before and after a settings change, while dVermin focuses on detecting text scaling issues. \revision{As SetDroid and dVermin rely on random testing, we used manual scripts to bypass the login functionality when needed, ensuring a fair evaluation.} Additionally, we evaluated GPT-4o, one of the state-of-the-art closed-source MLLMs, for bug detection using screenshots, following the methodology outlined in~\cite{liu2025seeingisbelieve, ju2024studyofusingmllm}.

Note that there are other state-of-the-art tools for detecting setting-related bugs, such as SetChecker~\cite{sun2023characterizing}, AccessiText~\cite{alshayban2022accessitext}, and ITDroid~\cite{escobar2020empirical}. These tools were excluded from our analysis for the following reasons. SetChecker was not included because it focuses on setting-related bugs caused by problematic Android API invocations, which do not align with our use case. ITDroid was excluded as it specifically targets setting-related bugs resulting from different language settings, which is not relevant to the scenario illustrated in Figure \ref{fig:motivating_example}. AccessiText was omitted because its code is unavailable.

\subsubsection{Results of Test Case Generation}

We first applied the above-mentioned baselines (a.k.a., SetDroid, dVermin, and GPT-4o) to the bug in Figure \ref{fig:motivating_example} to check whether they are able to trigger the bug. Specifically, we first checked whether the baselines can \textbf{generate valid test case to trigger the bug}. We then assessed if \textbf{the test oracles of the above approaches} could detect the SUD bug by triggering the bug and checking if a relevant report was generated.

We first checked the effectiveness of baselines in terms of test case generation process. Figure \ref{fig:reproduction_step} shows the reproduction steps for Signal issue \#11258. 
As we can see, even if app developers have set the system setting to dark mode, accessing the page that exhibits insufficient color contrast requires the app to encounter a runtime error, rather than triggering a UI action on the screen.

From the perspective of test case generation process, all the above approaches fail to actually trigger the bug depicted in Figure \ref{fig:motivating_example}. Specifically, for SetDroid and dVermin, their bug detection process is driven by random testing. Specifically, we allowed them to run for an hour, but both of them still could not trigger this bug because the random process failed to navigate to the archived-conversations page, which is required to reproduce the bug.

\subsubsection{Results of Test Oracle}
Regarding the test oracle, unfortunately, the aforementioned methods are still insufficient to detect this bug and produced false positives. 

First, for dVermin and ITDroid, their test oracles target bugs caused by text scaling rather than SUD bugs resulting from language changes. We now illustrate the results of SetDroid and GPT-4o as follows. 

Second, for SetDroid, its test oracle cannot be leveraged to detect the bug illustrated in Figure \ref{fig:motivating_example}. This is mainly because SetDroid's test oracle detects whether the UI state remains consistent by turning on and then off the specific system setting. However, this test oracle is not applicable for this type of SUD bug.

Third, for GPT-4o, although it can be leveraged to obtain the app semantics, it is unable to identify SUD bugs related to color contrast. GPT-4o produced a false warning by only reading the text content in the image, incorrectly indicating that ``an error has occurred on the Signal app.''

The above results show that existing state-of-the-art tools and approaches have limitations in detecting SUD bugs, particularly due to the lack of test oracles for covering the expected adaptations of system settings. To address these gaps, we introduce new test oracle concepts and implement them in SUDFinder, an automated Android UI testing tool for detecting SUD bugs. Instead of performing random testing, SUDFinder directly generates test cases for all XML configuration files of the tested UI page, in order to ensure that the generated test cases cover different UI pages in the app and their UI components. Regarding the test oracle, SUDFinder models common expected UI changes from various visual feature perspectives. For example, SUDFinder flagged the bug in Figure \ref{fig:motivating_example} because the text color changed from black to white, which should have been paired with a matching background change, but no such change occurred in dark mode. Modeling common expected UI changes requires knowledge derived from real-world SUD bugs. This motivates us to first conduct an empirical study on common patterns of SUD bugs, providing necessary guidance for the design of our automated approach.

\section{Empirical Study}
\label{sec:empirical}

To facilitate the automated detection of SUD bugs, we conducted an empirical study aimed at understanding the patterns of SUD bugs under different system settings. Specifically, the empirical study aims at answering the following Research Questions \textbf{(RQ)}:

\begin{itemize}
    \item \textbf{RQ1.} What are the common bug-triggering patterns of SUD bugs?
    \item \textbf{RQ2.} Where are SUD bugs commonly located in apps' code?
    \item \textbf{RQ3.} What are the consequences of SUD bugs?
    \item \revision{\textbf{RQ4.} Do app users consider SUD bugs important?}
\end{itemize}

\subsection{Empirical Study Setup}

\subsubsection{Setup for Empirical Dataset}

\begin{table}[t]
    \centering
    \caption{Setting Categories and Subcategories}
    \label{tab:SettingRelatedUIIssuesExtended}
    \begin{tabular}{>{\centering\arraybackslash}p{3cm}>{\centering\arraybackslash}p{5cm}}
        \toprule
        \textbf{Setting Category} & \textbf{Subcategory} \\
        \midrule
        \multirow{2}{*}{Screen (68)} & Rotation (53) \\
        & Multi-screen (15) \\
        \midrule
        \multirow{2}{*}{Scale (51)} & Display Size (14) \\
        & Text Size (37) \\
        \midrule
        \multirow{2}{*}{Language (35)} & LTR Language (16) \\
        & RTL Language (19) \\
        \midrule
        Theme (111) & Dark Mode (111) \\
        \midrule
        Others (43) & - \\
        \bottomrule
    \end{tabular}
\end{table}

To answer RQ1--3, we refer to the empirical dataset collected by Sun et al.~\cite{sun2021understanding, sun2023characterizing}, which includes 1,074 setting-related bugs from open-source Android apps. Among these, 218 bugs are categorized as SUD (Settings-UI-Display) bugs that directly affect the UI display of the apps. 
We have used these 218 SUD bugs as the baseline for our empirical study.

However, there is a threat that directly using the above empirical dataset (released in 2021) may not accurately reflect the current state of SUD bugs. To ensure up-to-date results, we repeated the dataset collection process described by Sun et al.~\cite{sun2021understanding, sun2023characterizing} to supplement our empirical dataset. The specific steps are as follows:

\textbf{Step 1: Subject App Selection.}
We selected open-source Android apps from GitHub as our study subjects to enable direct access to their source code, bug reports, fix patches, and developer discussions. These subjects were collected according to the following procedure.
\begin{itemize}[leftmargin=*]
    \item We used GitHub’s REST API to crawl all the Android projects on GitHub which are released on Google Play and F-Droid and have been actively maintained since 2021. We attained 937 Android projects.
    \item To ensure an adequate number of bug reports for our study, we applied a filter to include only projects with over 200 closed bug reports. We then attained 61 Android projects.
    \item We manually inspected each project, excluding those that are not real apps (e.g., demo projects for illustrating third-party libraries) and attained 58 Android projects.
\end{itemize}

\definecolor{headerbg}{RGB}{192,192,192} 
\definecolor{rowcolor1}{RGB}{245,245,245} 
\definecolor{rowcolor2}{RGB}{255,255,255} 

\begin{table*}[]
\centering
\caption{Keyword Sets Used for Filtering SUD Bug Reports ~\cite{sun2021understanding, sun2023characterizing}}
\label{tab:settings_keyword}
\begin{tabularx}{\linewidth}
{
    >{\hsize=0.5\hsize\arraybackslash}X
    >{\hsize=1.5\hsize\arraybackslash}X
}
\toprule
\textbf{Keyword Type} &
\textbf{Keywords} \\
\midrule
Setting & orientation, vertical, horizontal, split screen, Multi-window, screen insulation, brightness, landscape, portrait, rotate, developer option, keep activity, accessibility, talkback, text-to-speech, color correction, color inversion, high contrast text, setting, preference, date, time, time zone, hour format, date\&time, reading mode, car mode, one-handed mode, dark mode, game mode, night mode, theme, language \\
\midrule
UI-related     & preference, dark mode, night mode, theme, language, display, accessibility, text, color, contrast, orientation, vertical, horizontal, screen, resolution, brightness, portrait, landscape, rotate \\
\midrule
Defect/failure & crash, exception, bug, issue \\
\midrule
Reproducing    & repro, STR, record \\
\bottomrule
\end{tabularx}
\end{table*}

\textbf{Step 2: Keyword Search.}
From the 58 apps, we attained 85,479 bug reports in total. To identify relevant bug reports, we applied a keyword-based filtering strategy using four distinct sets of terms (summarized in Table \ref{tab:settings_keyword}). A bug report was selected for inclusion only if it contains at least one keyword from each of the four sets.
\begin{itemize}[leftmargin=*]
    \item \textit{Setting keywords}: While Sun et al.~\cite{sun2021understanding, sun2023characterizing} employed a broad set of keywords encompassing all system settings, we focus specifically on UI display-related settings (listed in Table \ref{tab:settings_keyword}) to align with our research objectives. For each keyword, we also consider the possible forms that users may use (e.g., capitalization, abbreviations, and tenses).
    \item \textit{UI-related keywords}: To ensure reports concern the rendered interface, we require each candidate to mention UI rendering or display-affecting behaviour via appearance/theming keywords (``preference,'' ``dark mode,'' ``night mode,'' ``theme,'' ``language,'' ``display,'' ``accessibility,'' ``text,'' ``color,'' ``contrast'') or screen/orientation keywords (``orientation,'' ``vertical,'' ``horizontal,'' ``screen,'' ``resolution,'' ``brightness,'' ``portrait,'' ``landscape,'' ``rotate''), excluding purely non-UI reports (e.g., background services or data sync).
    \item \textit{Defect/failure keywords}: We filter bug reports to isolate those describing actual defects or failures rather than feature requests or documentation bug reports. The filter uses the keywords ``crash,'' ``exception,'' ``bug,'' and ``issue.''
    \item \textit{Reproducing keywords}: We use the keywords of ``repro'', ``STR'', and ``record'' to filter bug reports that contain the reproducing steps, which are crucial for understanding and verifying whether a report describes a real, reproducible bug.
\end{itemize}

\revision{Applying the above keyword filter through GitHub's REST API initially yielded 6,076 bug reports. We then retrieved the full issue bodies, comments, and label metadata to verify that each report truly contains at least one keyword from each of the four sets. To retain only actionable bug reports, we further applied a label-based filter that excluded reports whose labels indicate: (1) rejected or non-defect reports (e.g., feature requests and enhancements); (2) redundant or invalid reports (e.g., duplicates and non-reproducible cases); (3) unresolved discussions or pending user input; (4) meta or non-substantive reports (e.g., documentation, and pull-request metadata); and (5) unconfirmed or untriaged reports. After this filtering, we obtained a final set of 1,247 bug reports within our study scope.}

\textbf{Step 3: Manual Inspection.}
Finally, we manually inspected the 1,247 candidate bug reports from the previous step, and kept only the valid bug reports according to the following rules: 
\begin{itemize}[leftmargin=*]
    \item We retained only the bug reports where the reporters or developers explicitly stated that changing system settings is a necessary condition for triggering the failures. Bug reports that just mention settings are excluded. 
    \item When the bug reports lacked definitive clues, we reproduced the failures by following their described steps to verify whether they reflect SUD defects.
\end{itemize}
We finally attained 90 valid bug reports, bringing the total size of the empirical dataset to 308 reports. Table \ref{tab:SettingRelatedUIIssuesExtended} presents the statistics of these SUD bugs and their associated settings. Notably, 86.0\% of the SUD bugs are related to screen, scale, language, and theme settings. 

The data analysis was conducted as follows. Initially, we took a random sample of 154 bug reports, representing 50\% of our empirical dataset. Two authors, each with two years of experience in Android app development, independently examined the code revisions and corresponding bug reports for each sampled bug to pinpoint code snippets associated with patches. A preliminary taxonomy was developed based on the findings of the two authors, with any disagreements resolved through meetings. The authors then iteratively labeled the remaining 154 bug reports, holding discussions to refine the preliminary taxonomy and resolve any conflicts. The final results were achieved once both authors reached an agreement on the taxonomy and the labels for the empirical dataset.

\subsubsection{Setup for RQ4}
\revision{To address RQ4, we conducted a user-centered study. In this study, we presented users with screenshots of SUD bugs and invited them to evaluate the impact of these defects on functionality and their user experience using a five-point Likert scale. We selected the top 15 projects with the highest star counts from our empirical research dataset and chose one bug report from each project. During the selection process, we first excluded bug reports that lacked relevant defect screenshots or did not provide reproduction steps. Then, for each project, we selected the most recently submitted bug report from the remaining candidates. Table \ref{tab:survey_result} shows the list of bug reports we ultimately selected. As shown in the table, these bug reports exhibit diversity in terms of 4 different types of setting categories.}

\revision{We conducted this survey on Amazon Mechanical Turk (AMT), a crowdsourcing platform widely used for user studies. Based on best practices from prior work, we implemented strict quality control over both the participants and their responses. We initially selected workers with a past task approval rate above 95\% and more than 5,000 approved tasks. We also excluded clearly inattentive responses, including: (1) more than 95\% of answers being identical, (2) answers following a specific pattern, and (3) page viewing times below the minimum limit (less than 60 seconds per question) or above the maximum limit (more than 600 seconds). There were 7 such cases in our collected sample. Finally, we collected 
60 valid survey responses.}


















\subsection{RQ1: Bug-Triggering Patterns}

In total, we identified and elaborated on the following triggering patterns of SUD bugs. 
Please note that since a report may contain multiple SUD bugs, the total percentage of the following categories will exceed 100\%.

\begin{figure}[t]
\centering
\includegraphics[width=0.48\textwidth]{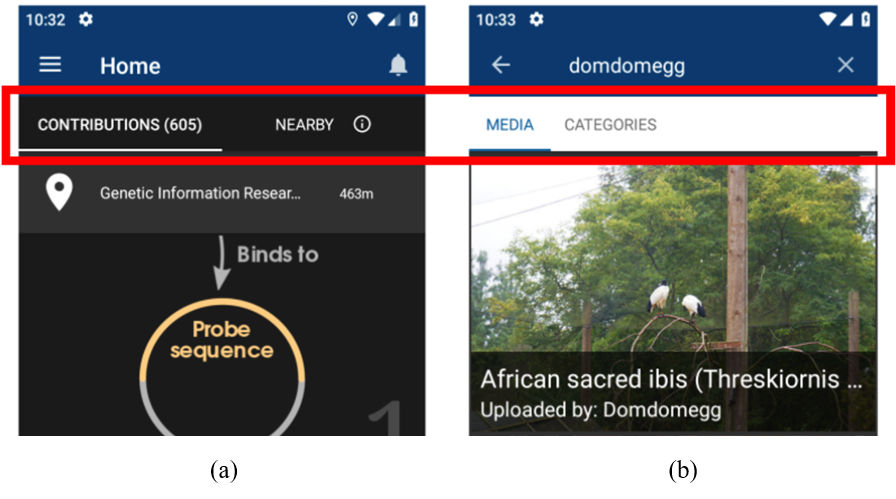}
\caption{An SUD bug reported in apps-android-commons \#2186~\cite{commons2186}, where the app developers failed to adjust the colors of toolbar in the dark mode as shown in (b)}
\label{fig:color_change}
\end{figure}

\begin{figure}[t]
\centering
\includegraphics[width=0.48\textwidth]{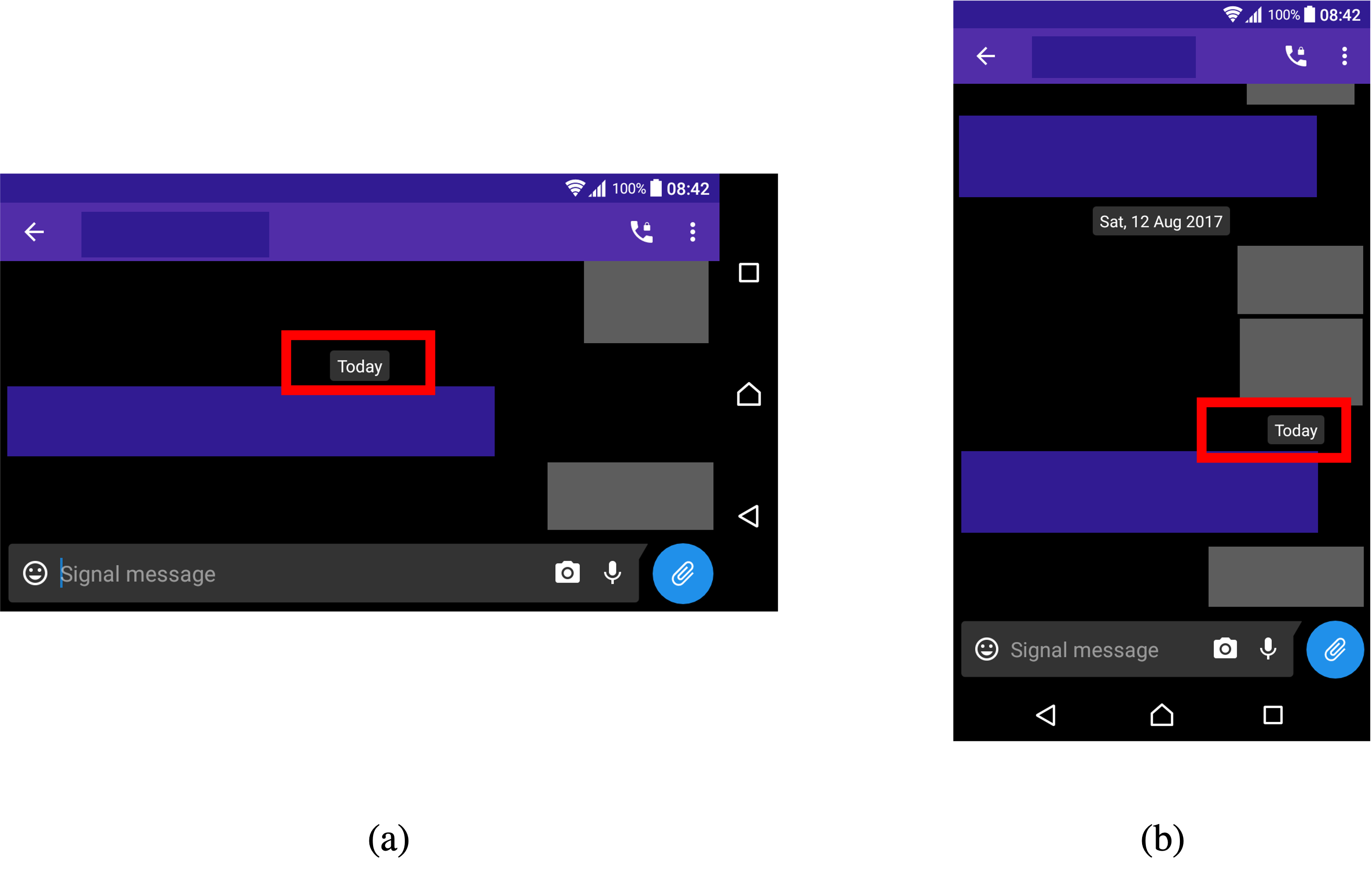}
\caption{An SUD bug reported in Signal-Android \#6875~\cite{signal6875}, where the ``Today'' label drifts off-center after rotation}
\label{fig:label_alignment_change}
\end{figure}

\begin{figure}[t]
\centering
\includegraphics[width=0.4\textwidth]{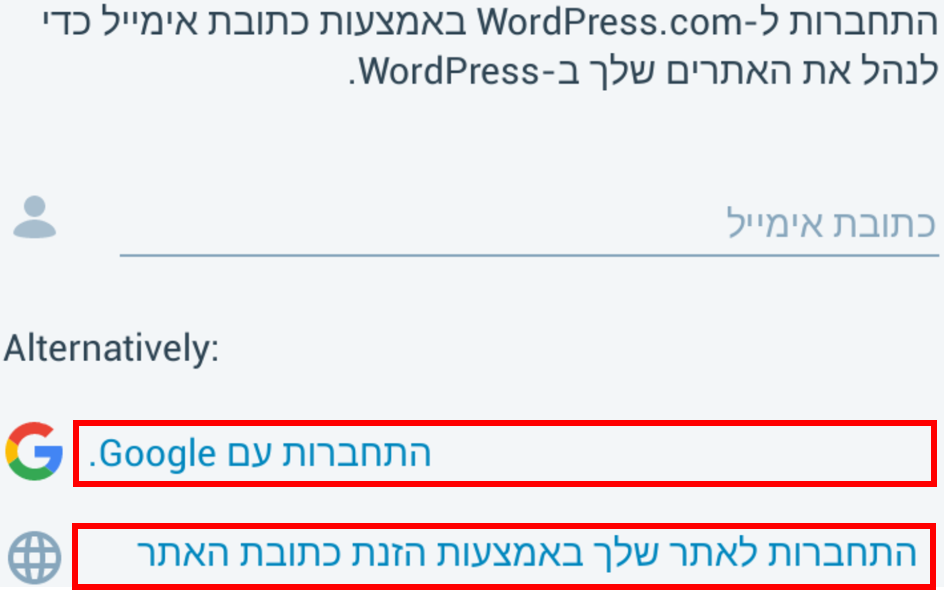}
\caption{An SUD bug reported in WordPress \#7324~\cite{wordpress7324}, where the texts fail to be aligned when changing the language to Hebrew.}
\label{fig:alignment_change}
\end{figure}

\begin{itemize}[leftmargin=1ex]

\item \textbf{T1: Style-Related SUD Bugs. }
We found that 222 SUD bugs are related to the UI styles (e.g., color, layout structures, spacing, etc.), affecting the consistency of UI pages under different system settings. We have identified the following subtypes for further clarification:

\begin{itemize}[leftmargin=1ex]
        \item \textbf{T1.1: Color Change Inconsistencies.} We found that 111 SUD bugs arise from inconsistencies in UI component colors. This type of bug can cause the app's UI components to have strange color schemes under different system settings, and sometimes even render them invisible to the users. As illustrated in Figure \ref{fig:color_change}(b), in apps-android-commons \#2186, developers neglected to update the toolbar colors in dark mode, resulting in a UI page that appeared inconsistent with other pages (see Figure \ref{fig:color_change}(a)).

        \item \textbf{T1.2: Alignment Inconsistencies.} A total of 63 SUD bugs occur when the position or size of a view changes, causing the visible content inside UI components to misalign under different system settings. This includes (1) 20 cases of alignment issues manifested as \textit{layout bound drift}, as shown in Figure~\ref{fig:label_alignment_change}, where the ``Today'' label drifts off-center after rotation, and (2) 43 cases where misalignment is not due to layout-bound drift, but rather caused by layout-bound internal rendering affecting visual content. An example is illustrated in WordPress \#7324, where text shifts left when the system language is set to Hebrew (see Figure~\ref{fig:alignment_change}). 
        
        \item \textbf{T1.3: Distance Inconsistencies.} Developers commonly place semantically-related UI components close together for visual coherence. The expected distance between these components should only change under specific system settings related to screen size, such as screen rotation. However, changes in other system settings can cause these components to become distant, leading to SUD bugs. In our empirical dataset, we identified 36 such bugs. Note that while T1.3 focuses on distance changes between UI components, alignment bugs (T1.2) can also indirectly cause changes in the distances between components. We identified 11 distance inconsistency bugs related to this.

        \item \textbf{T1.4: Containment Inconsistencies.} Another type of SUD bugs occurs when the containment (i.e., the contains and overlaps relationships) between two UI components is inconsistent across different system settings. For instance, a container view may not adjust its size or position to properly encapsulate a contained view during dynamic changes. There are 34 cases in the empirical dataset.

    \end{itemize}

    The remaining 2 SUD bugs affect other UI styles, including font, text shadows, etc.

    \item \textbf{T2: Semantic-Related SUD bugs. }We found 86 SUD bugs related to the semantics of UI components that did not correctly follow system setting changes, causing functional failure. The subtypes are distilled as follows:

\begin{itemize}[leftmargin=1ex]
    \item \textbf{T2.1: Unexpected UI Visibility.} There are 48 cases where UI components fail to update correctly, resulting in persistent notifications or dialogs that misrepresent the app's current state. For example, in open-event-attendee-android \#1551~\cite{openevent1551}, the ``No Internet'' dialog remains visible even after users enable their network connection.

    \item \textbf{T2.2: Improper Data Handling.} There are 19 bugs when app developers fail to properly manage the data displayed in UI components (e.g., data loss), which can even lead to app crashes. For instance, in focus-android \#3167~\cite{focus3167}, the app crashes because the developers did not account for the scenario where network data cannot be fetched while in airplane mode.

    \item \textbf{T2.3: Translation Inconsistencies. } We identified 19 SUD bugs related to string translations failing to meet the specification of system setting changes. These bugs manifested as incorrect linguistic formatting (e.g., erroneous word order) as well as inaccurate or incomplete translations.  
\end{itemize}
    
\end{itemize}

\begin{figure}[t]
\centering
\includegraphics[width=0.48\textwidth]{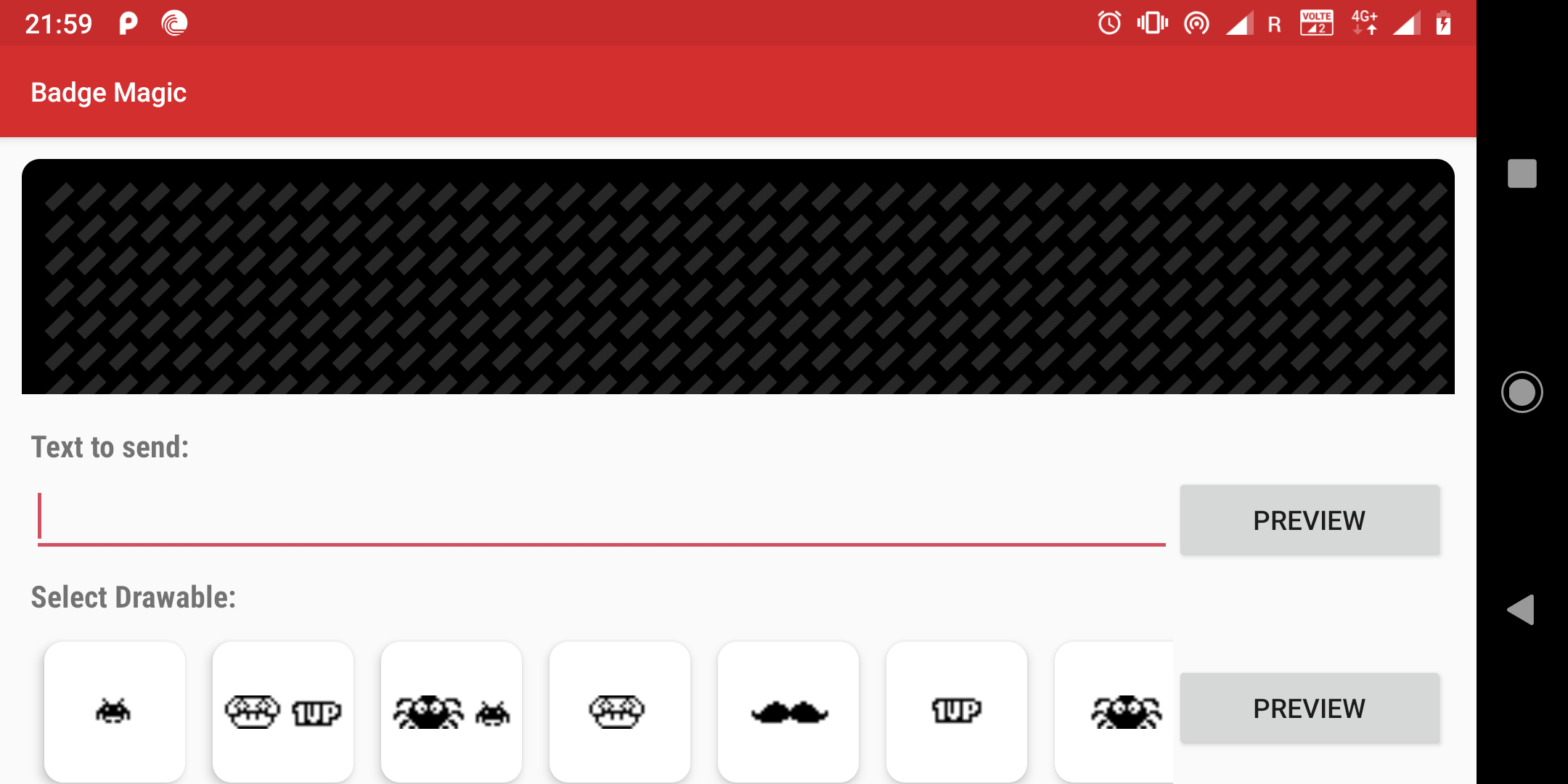}
\caption{An SUD bug reported in badgemagic-android \#106~\cite{badgemagic106}, where the icon becomes partially visible when rotating the screen.}
\label{fig:visibility_change}
\end{figure}

\subsection{RQ2: Bug Locations}

We further analyzed the locations where these SUD bugs occur. There are 188 bugs that occur in XML configuration. 
The bug in Figure 1 is such a case because app developers fail to adapt the text color to the dark theme. Of these configuration bugs, 154 were categorized as T1-type and 34 as T2-type.
Furthermore, 120 bugs originated from the app code, especially due to developers not dynamically adapting the visual appearance of UI components to different system settings. For example, a certain bug might cause inconsistent look-and-feel across different API levels, affecting the app's functionalities. Within this group, 68 were T1-type and 52 were T2-type. The above findings motivate the design of our bug detection approach to conduct analysis at the level of XML files.

\subsection{RQ3: Bug Consequences}

SUD bugs can lead to inconsistent app behavior across different system settings, manifesting as either functional or non-functional issues. Below, we discuss some common symptoms.

Of the 308 SUD bugs in the dataset, 168 are functional, directly preventing users from effectively utilizing the app's features. For instance, as illustrated in Figure \ref{fig:visibility_change}, one bug renders an icon unclickable due to an overlap between the icon area and the preview button.
Notably, only 22 out of the 308 bugs resulted in an app crash. This indicates that the implicit oracles commonly employed in test automation are insufficient for detecting the remaining 286 functional and non-functional issues. Defining, implementing, and validating robust test oracles for non-crashing bugs remains one of the most challenging aspects of test automation, underscoring the importance of  research towards bug classification, which serve as the guidance towards the design of SUDFinder.

The remaining 140 SUD bugs are non-functional, causing UI variations across different system settings and potentially affecting the user experience. One such example is shown in Figure \ref{fig:color_change}, where the snack bar’s color does not align with the dark mode color scheme.

These findings underscore the severity of SUD bugs and motivate the need for an automated approach to detect them effectively.

\subsection{RQ4: Bug Severity}
\revision{The results of the user study are presented in Table \ref{tab:survey_result}. From a functionality perspective, on average 37.2 out of 60 participants (62.0\%) awarded a score of 4 or higher, with an average of 12.6 participants (21.0\%) giving the highest score of 5. Notably, issues related to internationalization/localization and landscape mode were identified as having functional shortcomings. These bugs were primarily attributed to text truncation or content occlusion, which prevented users from viewing all relevant information. Furthermore, from the perspective of user experience, on average 35.7 out of 60 participants (59.4\%) gave a score of 4 or above, with an average of 15.3 participants (25.6\%) awarding a score of 5. These findings suggest that users generally perceive SUD bugs as having a non-trivial impact on both functionality and user experience.}

\begin{table*}[!htb]
\centering
\ifshowrevisioncomments\color{blue}\fi
\caption{Survey Result}\label{tab:survey_result}
\resizebox{\textwidth}{!}{
\begin{threeparttable}
\begin{tabular}{cccccccccccccccc}

\toprule
\multirow{2}{*}{\textbf{Issue}} & \multirow{2}{*}{\textbf{Category}} & \multicolumn{7}{c}{\textbf{Functionality}} & \multicolumn{7}{c}{\textbf{User Experience}} \\
\cmidrule(lr){3-9} \cmidrule(lr){10-16}
 &  & \textbf{5} & \textbf{4} & \textbf{3} & \textbf{2} & \textbf{1} & \textbf{5+4}\tnote{*} & \textbf{Avg} & \textbf{5} & \textbf{4} & \textbf{3} & \textbf{2} & \textbf{1} & \textbf{5+4}\tnote{*} & \textbf{Avg}\\
 \midrule
Signal-Android \#6265  & Dark Mode & 21.7\% & 46.7\% & 26.7\% & 3.3\% & 1.7\% & 68.3\% & 3.8 & 33.3\% & 21.7\% & 21.7\% & 18.3\% & 5.0\% & 55.0\% & 3.6\\
Seal \#1075 & Large Font/Display Size & 23.3\% & 41.7\% & 20.0\% & 6.7\% & 8.3\% & 65.0\% & 3.6 & 31.7\% & 30.0\% & 20.0\% & 15.0\% & 3.3\% & 61.7\% & 3.7\\
thunderbird-android \#1930  & Dark Mode & 20.0\% & 31.7\% & 33.3\% & 13.3\% & 1.7\% & 51.7\% & 3.5 & 18.3\% & 38.3\% & 20.0\% & 21.7\% & 1.7\% & 56.7\% & 3.5\\
Aegis \#754  & Dark Mode & 23.3\% & 35.0\% & 21.7\% & 13.3\% & 6.7\% & 58.3\% & 3.5 & 30.0\% & 28.3\% & 20.0\% & 11.7\% & 10.0\% & 58.3\% & 3.6\\
StreetComplete \#1372 & Language & 15.0\% & 40.0\% & 38.3\% & 3.3\% & 3.3\% & 55.0\% & 3.6 & 21.7\% & 40.0\% & 20.0\% & 15.0\% & 3.3\% & 61.7\% & 3.6\\
ReadYou \#685  & Large Font/Display Size & 18.3\% & 46.7\% & 15.0\% & 18.3\% & 1.7\% & 65.0\% & 3.6 & 26.7\% & 25.0\% & 26.7\% & 16.7\% & 5.0\% & 51.7\% & 3.5\\
AmazeFileManager \#4225 & Language & 15.0\% & 41.7\% & 35.0\% & 8.3\% & 0.0\% & 56.7\% & 3.6 & 10.0\% & 41.7\% & 31.7\% & 16.7\% & 0.0\% & 51.7\% & 3.5\\
nextcloud \#14717  & Language & 30.0\% & 40.0\% & 16.7\% & 8.3\% & 5.0\% & 70.0\% & 3.8 & 26.7\% & 35.0\% & 26.7\% & 8.3\% & 3.3\% & 61.7\% & 3.7\\
cgeo \#14203  & Landscape Mode & 26.7\% & 33.3\% & 26.7\% & 8.3\% & 5.0\% & 60.0\% & 3.7 & 26.7\% & 33.3\% & 25.0\% & 5.0\% & 10.0\% & 60.0\% & 3.6\\
status-im \#5025 & Landscape Mode & 18.3\% & 35.0\% & 35.0\% & 11.7\% & 0.0\% & 53.3\% & 3.6 & 30.0\% & 38.3\% & 15.0\% & 10.0\% & 6.7\% & 68.3\% & 3.8\\
AnySoftKeyboard \#3001  & Dark Mode & 20.0\% & 43.3\% & 25.0\% & 3.3\% & 8.3\% & 63.3\% & 3.6 & 21.7\% & 35.0\% & 23.3\% & 11.7\% & 8.3\% & 56.7\% & 3.5\\
attendee \#960 & Landscape Mode & 20.0\% & 50.0\% & 18.3\% & 11.7\% & 0.0\% & 70.0\% & 3.8 & 26.7\% & 33.3\% & 28.3\% & 11.7\% & 0.0\% & 60.0\% & 3.8\\
Gramophone \#213  & Large Font/Display Size & 16.7\% & 48.3\% & 20.0\% & 11.7\% & 3.3\% & 65.0\% & 3.6 & 31.7\% & 41.7\% & 15.0\% & 6.7\% & 5.0\% & 73.3\% & 3.9\\
badgemagic-android \#106 & Landscape Mode & 23.3\% & 46.7\% & 15.0\% & 13.3\% & 1.7\% & 70.0\% & 3.8 & 21.7\% & 33.3\% & 30.0\% & 8.3\% & 6.7\% & 55.0\% & 3.5\\
kiwix-android \#1842 & Landscape Mode & 23.3\% & 35.0\% & 31.7\% & 8.3\% & 1.7\% & 58.3\% & 3.7 & 26.7\% & 33.3\% & 25.0\% & 10.0\% & 5.0\% & 60.0\% & 3.7\\
\midrule
\multicolumn{2}{c}{\textbf{Average}} & \textbf{21.0\%} & \textbf{41.0\%} & \textbf{25.2\%} & \textbf{9.6\%} & \textbf{3.2\%} & \textbf{62.0\%} & \textbf{3.7} & \textbf{25.6\%} & \textbf{33.9\%} & \textbf{23.2\%} & \textbf{12.4\%} & \textbf{4.9\%} & \textbf{59.4\%} & \textbf{3.6}
\\
 \bottomrule
 \end{tabular}
\begin{tablenotes}
\footnotesize
\item[*] ``5+4'' denotes the percentage of participants who rated the bug report 5 or 4 on the five-point Likert scale (i.e., the combined proportion of the two highest ratings).
\end{tablenotes}
\end{threeparttable}

}
\end{table*}

\subsection{Empirical Findings Summary}

Note that Sun et al.~\cite{sun2021understanding, sun2023characterizing} proposed the above dataset and identified common bug patterns related to settings. Their empirical study mentioned SUD bugs that may affect the apps' UI but did not provide a detailed classification of SUD bugs. Additionally, the common bug patterns related to settings that they identified have been integrated into the automated tools SetDroid and SetChecker. This bug detection approach operates under the assumption that app behaviors should remain consistent if a given setting is changed and later properly restored, or show expected differences if not restored. Nevertheless, the above proposed test oracle is effective for 16.5\% (36/218) of SUD bugs in the empirical dataset of Sun et al. \cite{sun2021understanding, sun2023characterizing}, as they are commonly manifested at the buggy system setting and remain consistent in the default setting. The SUD bug illustrated in Figure \ref{fig:motivating_example} shows such an example, as the bug only manifests when changing to the dark mode and behaves consistently when changing back to the default setting.

Existing studies~\cite{escobar2020empirical, su2022metamorphosis, alshayban2022accessitext, chen2021accessible} have summarized SUD bugs caused by specific types of settings. For example, Su et al.\cite{su2022metamorphosis} and Alshayban et al.\cite{alshayban2022accessitext} summarized common patterns of SUD bugs resulting from text scaling adjustments, while Escobar et al.~\cite{escobar2020empirical} focused on SUD bugs related to language changes. Although the aforementioned methods can target at style-related UI bugs, they only cover T1.2 and T1.4 bugs, failing to handle T1.1 and T1.3 SUD bugs, which account for 66.2\% (147/222) 
of the T1 SUD bugs. Additionally, regarding T1.2 SUD bugs, existing work~\cite{escobar2020empirical} only considers alignment related to layout bounds, without analyzing the alignment that requires examination of the content rendered on the UI page (e.g., WordPress \#7324). This overlooked category constitutes a majority of T1.2 bugs in the empirical dataset, with 43 out of 63 instances (68.2\%).

\section{Approach}
\label{sec:approach}
\subsection{Overview}
In this paper, we propose SUDFinder to automatically detect SUD bugs. SUDFinder primarily focuses on T1 bugs, as they account for 72.1\% 
in the empirical dataset. Although existing approaches have been proposed to address layout containment inconsistencies (34 T1.4 bugs)~\cite{su2022metamorphosis, mahajan2015detection, alshayban2022accessitext} and layout alignment inconsistencies related to layout bounds (20 T1.2 bugs)~\cite{escobar2020empirical, mahajan2015detection}, these approaches lack a comprehensive, general-purpose solution for T1 bugs. \revision{We do not cover T2 bugs: T2.2 (19 bugs) and T2.3 (19 bugs) are already addressed by existing data-loss detectors~\cite{guo2022detecting, zhou2023ddldroid, riganelli2020data} and ITDroid~\cite{escobar2020empirical} respectively, so we do not re-implement them; T2.1 (48 bugs) requires reasoning over UI semantics across user-event sequences, a technical direction orthogonal to our XML-structure-based analysis, and given its small proportion (48/308) we leave it as future work.}

To detect T1 bugs, the intuition behind SUDFinder is that \textit{UI components with similar visual features should undergo the same adaptation when settings change, thereby maintaining a coherent visual effect after system setting changes}. The example in Figure \ref{fig:motivating_example} illustrates this: when the system setting switches to the dark mode, the text color should be changed accordingly in order to maintain the readability. This motivates us to adopt different metrics to assess the relations between UI components on the page, identifying those faulty UI components that fail to maintain these relations and generating bug reports.

\begin{figure*}[t]
\centering
\includegraphics[width=0.85\textwidth]{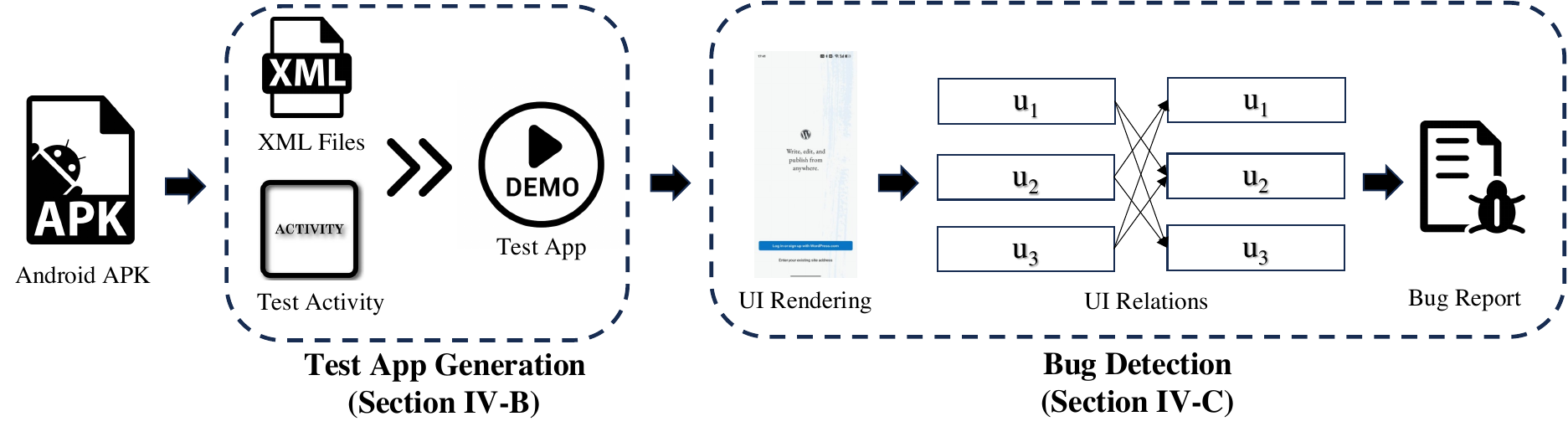}
\caption{Overview of SUDFinder.}
\label{fig:overview}
\end{figure*}

Figure \ref{fig:overview} provides an overview of SUDFinder. Specifically, given the APK file of an Android app as input, SUDFinder first extracts all XML configuration files related to the app’s UI from the APK, and generates a test app for each XML configuration file. Each test app contains a test activity that can display the visual appearance of the corresponding XML configuration file (See Section \ref{sec:testappgeneration}). Then, SUDFinder dynamically runs these test apps and analyzes the relationships of the UI components declared in the XML configuration files under different system settings. If SUDFinder detects a UI relationship that violates the specification of system setting changes, a bug report is generated (See Section \ref{sec:bugdetectionprocess}).

\begin{figure}[t]
\centering
\includegraphics[width=0.48\textwidth]{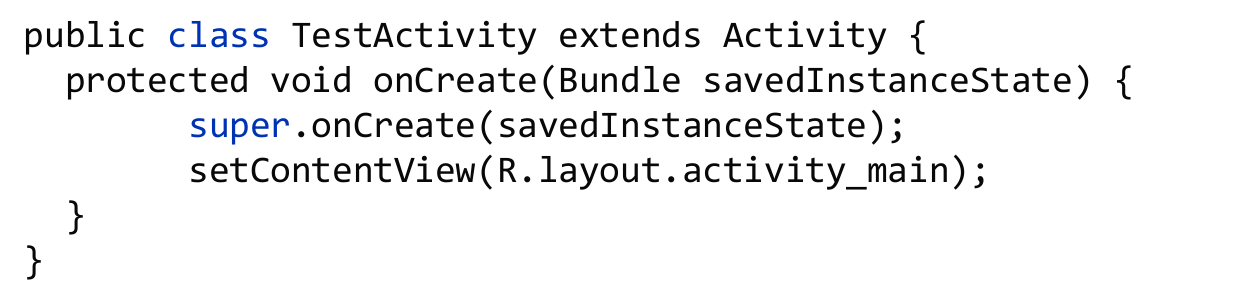}
\caption{A test activity for Signal issue \#11258~\cite{signal11258}}
\label{fig:inject_activity}
\end{figure}

\subsection{Test App Generation}
\label{sec:testappgeneration}
Initially, SUDFinder injects a test activity into $app$ to reveal the visual appearance of all the UI pages. Figure \ref{fig:inject_activity} illustrates an example of the test activity injected to detect the bug shown in Figure \ref{fig:motivating_example}. Specifically, the test activity includes the invocation of \texttt{setContentView} by passing the ID of the UI page as the parameter. We made this design choice because XML configuration files are widely used to build apps' UI, and many SUD bugs are located in XML configuration files (see Section~\ref{sec:empirical}). By doing this, one can automate the detection of SUD bugs and achieve high coverage on apps' UI without the need to build dynamic test cases to actually explore the UI components. During the execution of a test app, SUDFinder checks if the currently visible screen is capable of horizontal or vertical scrolling. SUDFinder performs a scroll action and retrieves any additional UI elements that appear afterwards. This step is crucial for revealing views that were initially part of the screen layout under the original settings (e.g., default font size) but have been pushed out of view (e.g., large font size). The resulting list of newly visible views allows for a comprehensive comparison of all displayed views, both with and without system setting changes.

In practice, since the UI pages in the app often contain text fetched from the web rather than loaded from predefined strings within the app, the absence of such text can render the TextView’s UI ineffective. As shown in Figure \ref{fig:patch}, the content of the TextView with the ID \texttt{reminder\_text} is not specified in the XML configuration file but is fetched from the server at runtime. Many of these TextViews lack test cases, presenting a challenge in finding text that aligns with actual application scenarios.

To tackle this challenge, SUDFinder leverages MLLMs to generate the texts missed in the XML configuration files. This approach is particularly effective because MLLMs are trained on extensive corpora from the internet, enabling them to acquire significant domain knowledge. In this paper, we provide the MLLM with the context of the UI component under test, allowing it to infer suitable text for the TextView. This method aims to generate realistic and contextually appropriate text for testing purposes. By doing so, we enhance the accuracy of UI rendering and help identify potential issues arising from different text inputs.

\begin{figure}[t]
\centering
\includegraphics[width=0.45\textwidth]{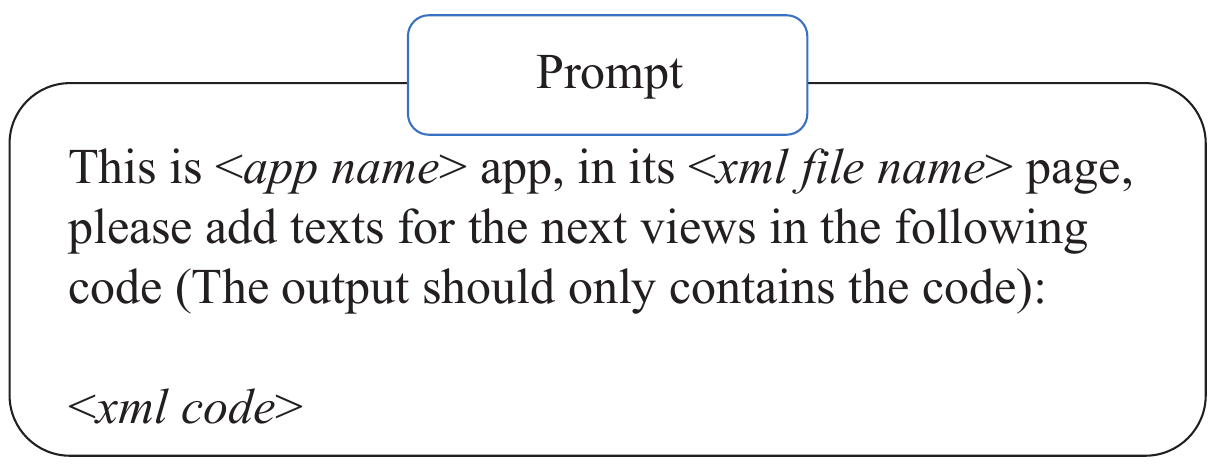}
\caption{Prompt template for inferring potential texts in TextViews or EditTexts.}
\label{fig:prompt_template}
\end{figure}

Figure \ref{fig:prompt_template} shows the prompt template of SUDFinder. Specifically, the template takes the app name, XML configuration file name, and XML code as inputs, allowing the MLLM to infer the context of UI pages in the XML code to generate appropriate text content. We ensure the MLLM outputs only the XML code without additional information, using the generated XML to replace the old file to build the test apps.

There are potential threats that the process may not accurately recover the visual appearance of the UI page under test, leading to false positives and negatives. Currently, we have not identified any false positives. False negatives can occur when SUDFinder fails to recover the visual appearance of a buggy UI page in both the baseline and tested settings. Although we cannot cover all cases of system setting changes, the evaluation results show that test activity injection helps detect 35 more SUD bugs compared to the random exploration strategy established by Sun et al.\cite{sun2021understanding, sun2023characterizing} (randomly choose an executable widget from the UI layout and generate an event), and 24 SUD bugs are located in the texts generated by MLLM (see Section \ref{sec:evaluation}).

\subsection{Bug Detection Process}
\label{sec:bugdetectionprocess}

\begin{algorithm}[t]
    \caption{SUDFinder Overview}
    \KwIn{App under test \textit{app}}
    \KwOut{Bug reports $BUGS$}
    
    Initialize $BUGS \leftarrow \emptyset$\;
    
    \ForEach{$page \in AUT$}{
        \ForEach{$(u_1, u_2) \in page$}{
            Obtain $R(u_1, u_2)$ under default setting\;
            Obtain $R(u_1', u_2')$ after setting change\;
            \If{buggyRelation($R(u_1, u_2), R(u_1', u_2')$)}{
                Add the bug to $BUGS$\;
            }
        }
    }
    \Return $BUGS$\;
    \label{alg:overview}
\end{algorithm}

Algorithm 1 provides an overview of SUDFinder. For each UI page $page$ in the Android app $app$, SUDFinder first extracts the visual features of UI components $U$ in $page$. Formally, a UI component $u \in U$ is defined as a tuple of $<bnd_{vh}, bnd_{vis}, color>$. Specifically:

\begin{itemize}[leftmargin=*]

\item $bnd_{vh}$ denotes the layout boundary of $u$ in the view hierarchy 
, represented as $<(x_{vh}^{l}, y_{vh}^{l}), (x_{vh}^{r}, y_{vh}^{r})>$. Here, $(x_{vh}^{l}, y_{vh}^{l})$ are the coordinates of the top-left corner, and $(x_{vh}^{r}, y_{vh}^{r})$ are those of the bottom-right corner. 

\item $bnd_{vis}$ denotes the Minimum Bounding Rectangle (MBR) that can cover all non-transparent pixels within $bnd_{vh}$, representing the visual content in $u$ 
. It is represented as $<(x_{vis}^{l}, y_{vis}^{l}), (x_{vis}^{r}, y_{vis}^{r})>$, where $(x_{vis}^{l}, y_{vis}^{l})$ are the coordinates of the top-left corner, and $(x_{vis}^{r}, y_{vis}^{r})$ are those of the bottom-right corner.

\item $color$ refers to the two most dominant colors displayed on $u$, defined as a tuple $<c_1, c_2>$. Following the practice of \cite{zhang2023automated}, SUDFinder determines this by invoking the \texttt{\#getcolors()} function from the Python Image Library~\cite{getcolors}. Although a rendered UI component may contain multiple colors, this design is based on the observation that the color composition of a single component is relatively simple, usually involving just the background color and the color of the text or image on it~\cite{zhang2023automated}.

\end{itemize}

Then, for any two UI components $u_1$ and $u_2$ from $page$ (Line 3), SUDFinder extracts the relation between $u_1$ and $u_2$ under the default setting (denoted as $R(u_1, u_2)$) and the relation $R(u_1', u_2')$ after a setting change (Line 4--5). SUDFinder generates a bug report if the relations $R(u_1, u_2) \neq R(u_1', u_2')$ (Line 6--7).

Based on the empirical findings in Section \ref{sec:empirical}, SUDFinder typically monitors the following four visual feature relations $R(u_1,u_2)$ for any pairs of UI components $u_1$ and $u_2$ as follows: 

\subsubsection{Color Change Inconsistencies} 
To detect T1.1 bugs, SUDFinder assumes that two UI components with the same colors under default settings should maintain consistent colors after a setting change. Therefore, SUDFinder generates a bug report if it finds that:
\[
u_1.color = u_2.color \quad \text{and} \quad u_1'.color \neq u_2'.color
\]

\subsubsection{Alignment Change Inconsistencies} 
To detect T1.2 bugs, SUDFinder assumes that two UI components, \( u_1 \) and \( u_2 \), which are aligned under default settings, should remain aligned after a settings change. Specifically, SUDFinder compares the visual bounds \( bnd_{vis} \) of \( u_1 \) and \( u_2 \) at six key points: left (L), right (R), top (T), bottom (B), vertical center (VC), and horizontal center (HC), as outlined in Algorithm 2. We define \( u_1 \) and \( u_2 \) as aligned if:
\[
u_1.align \cap u_2.align \neq \emptyset
\]

SUDFinder generates a bug report if, after a system change, it detects that:
\[
u_1'.align \cap u_2'.align = \emptyset
\]

\subsubsection{Distance Inconsistencies}
To detect T1.3 bugs, SUDFinder assumes that, unless system setting changes lead to screen alterations (e.g., screen rotation), the distance between the visual content of two UI components should remain as consistent as possible with the default settings. Therefore, SUDFinder calculates the horizontal Euclidean distance \( d_h(u_1, u_2) \) and the vertical Euclidean distance \( d_v(u_1, u_2) \) between the center points of the visible boundary $bnd_{vis}$ of components \( u_1 \) and \( u_2 \). SUDFinder generates a bug report if the changes in the horizontal and vertical Euclidean distances before and after the system setting changes exceed the thresholds in horizontal \( t_h \) and vertical \( t_v \) as shown below:
\[
 |d_h(u_1', u_2')-d_h(u_1, u_2)| \geq t_h \text{ or }  |d_v(u_1', u_2')-d_v(u_1, u_2)| \geq t_v
\]

\subsubsection{Overlap Inconsistencies}
To detect T1.4 bugs, SUDFinder checks the overlap of $bnd_{vis}$ between $u_1$ and $u_2$ before and after a system setting change (denoted as $overlap(u_1, u_2)$). It reports a bug if:
\[
 overlap(u_1, u_2) \neq overlap(u_1', u_2')
\]

\begin{algorithm}[t]
\label{alg:alignment}\small
\caption{Alignment Detection}
\KwIn{UI components $u_1, u_2$}
\KwOut{Alignment array $align$}

$align \leftarrow \emptyset$;

$u_1.x_{vis}^c \leftarrow \frac{u_1.x_{vis}^l + u_1.x_{vis}^r}{2}$, 
$u_2.x_{vis}^c \leftarrow \frac{u_2.x_{vis}^l + u_2.x_{vis}^r}{2}$, 
$u_1.y_{vis}^c \leftarrow \frac{u_1.y_{vis}^l + u_1.y_{vis}^r}{2}$, 
$u_2.y_{vis}^c \leftarrow \frac{u_2.y_{vis}^l + u_2.y_{vis}^r}{2}$

\textbf{if} $u_1.x_{vis}^l = u_2.x_{vis}^l$ \textbf{then} $align \gets align \cup \{\text{$L$}\}$

\textbf{if} $u_1.x_{vis}^r = u_2.x_{vis}^r$ \textbf{then} $align \gets align \cup \{\text{$R$}\}$

\textbf{if} $u_1.x_{vis}^c = u_2.x_{vis}^c$ \textbf{then} $align \gets align \cup \{\text{$HC$}\}$

\textbf{if} $u_1.y_{vis}^l = u_2.y_{vis}^l$ \textbf{then} $align \gets align \cup \{\text{$T$}\}$

\textbf{if} $u_1.y_{vis}^r = u_2.y_{vis}^r$ \textbf{then} $align \gets align \cup \{\text{$B$}\}$

\textbf{if} $u_1.y_{vis}^c = u_2.y_{vis}^c$ \textbf{then} $align \gets align \cup \{\text{$VC$}\}$

\Return $align$\;

\end{algorithm}

\section{Evaluation}
\label{sec:evaluation}
We implemented SUDFinder and evaluated it on real-world apps to answer the following research questions:
\begin{itemize}[leftmargin=*]
\item \textbf{RQ5.} How effective is SUDFinder in detecting SUD bugs?
\item \revision{\textbf{RQ6.} How do the MLLM module and test app generation strategy each contribute to bug detection?}
\item \textbf{RQ7.} How does the test oracle contribute to the bug detection?
\item \textbf{RQ8.} Can MLLMs be directly used to detect SUD bugs?
\end{itemize}

\subsection{Evaluation Setup}

\subsubsection{Evaluation Subjects}
We conducted the experiments on 29 open-source Android apps extracted from F-Droid~\cite{fdroid}, a well-known repository for such apps, as detailed in Table \ref{tab:evaluationsubjects}. In selecting these apps, we established the following criteria: (1) the app has more than 2,000 stars on GitHub~\cite{github}, (2) the app has had commits in recent months at the time of our experiments, and (3) the app is open-source to ensure the reproducibility of results. These criteria aim to filter for popular and actively maintained Android apps, ensuring that the bugs we identify have a significant impact. Using the above criteria, we selected 193 apps from the 4,037 available on F-Droid. We further excluded any apps that met one or more of the following conditions: (1) the app was included in the empirical study dataset~\cite{sun2021understanding,sun2023characterizing} to avoid overfitting our evaluation results; (2) the app cannot be installed and run on Android emulators; (3) the project is a game app; (4) the app is a toy app (e.g., proof-of-concept apps) or a duplicate of others; (5) the app cannot run after the test activity injection. Following these guidelines, we selected 29 apps as our experimental subjects. These apps are diverse (covering 13 different categories) and widely used with thousands or millions of downloads.

\begin{table}[t]
\centering
\caption{Evaluation Subjects}
\label{tab:evaluationsubjects}
\resizebox{0.48\textwidth}{!}{
\begin{tabular}{cccc}
\toprule
\textbf{App Name} & \textbf{Category} & \textbf{Stars} & \textbf{Downloads} \\
\midrule
Anki-Android & Education & 8.2K & 10M+ \\
afwall & Productivity & 3.2K & 500K+ \\
AntennaPod & Music \& Audio & 6.1K & 1M+ \\
AmazeFileManager & Tools & 5.2K & 1M+ \\
AppManager & - & 4.6K & - \\
breezy-weather & - & 4.7K & - \\
connectbot & Communication & 2.4K & 5M+ \\
droidify & - & 3.3K & - \\
InnerTune & - & 3.7K & - \\
KeePassDX & Productivity & 4.4K & 100K+ \\
KISS & Personalization & 2.9K & 100K+ \\
LibreTube & - & 8.3K & - \\
MaterialFiles & Tools & 5.6K & 1M+ \\
NewPipe & - & 30K & - \\
Omni-Notes & Productivity & 2.7K & 1K+ \\
osmdroid & Map & 2.8K & 10K+ \\
owncloud & Productivity & 3.8K & 100K+ \\
Password Store & Security & 2.5K & 10K+ \\
plain-app & Productivity & 2.4K & 100K+ \\
ReadYou & - & 4.5K & - \\
rethink-app & Finance & 2.7K & 100K+ \\
RetroMusicPlayer & Music & 3.9K & 1M+ \\
Simple-Calendar & Productivity & 3.5K & 10M+ \\
spotiflyer & - & 10K & - \\
tasks & News & 3.2K & 3M+ \\
Tusky & Social & 2.4K & 500K+ \\
uhabits-dev & Productivity & 7.6K & 5M+ \\
ViMusic & Entertainment & 8.3K & 100+ \\
WifiAnalyzer & Tools & 2.9K & 10M+ \\
\bottomrule
\end{tabular}
}
\end{table}

\subsubsection{Runtime Environment}

Our experiments were conducted on a server equipped with a 64-core Intel Xeon CPU E5-2683 v4 @ 2.10GHz processor and 192GB of RAM. On this hardware, we ran an Android emulator (Pixel 7) configured with API level 35, 16GB RAM, 64GB ROM, and a display resolution of 1440×2960 pixels. We leveraged GPT-4o to conduct text generation during the test app generation process. To determine horizontal and vertical distance inconsistencies, we follow the practice of Fazzini et al.~\cite{fazzini2017automated} by setting thresholds \( t_h \) and \( t_v \) to 10\% of the screen's width and height (144 and 296 pixels, respectively). These thresholds are used to identify horizontal and vertical distance inconsistencies. We followed \cite{su2022metamorphosis} to evaluate SUDFinder under 3 different scale settings: default display size and default font size, largest display size and default font size, largest display size and largest font size. We use the same evaluation setup as ITDroid except for the language settings. Specifically, we used English as the default language, and for the target languages we used Spanish, Hindi, and Arabic, because this set of languages is based on the list of languages with the largest number of speakers, and the IBM Translation Service leveraged by ITDroid does provide translations from English to those languages~\cite{escobar2020empirical}. The selected languages cover both LTR and RTL languages.

\subsection{RQ5: \revision{SUDFinder Bug Detection Effectiveness}}
\subsubsection{Setup}
We run SUDFinder on the selected evaluation subjects. For each bug report generated by SUDFinder, we conducted manual reproduction trying to validate the warnings reported by SUDFinder and baselines, following the practices of Huang et al.~\cite{huang2021characterizing}. Specifically, we first tried to build test cases to reach the \texttt{Activity} using the XML element in the bug reports. Then, we tried to manifest the visual effect controlled by the buggy UI components by reading the code logic of the \texttt{Activity}. A bug report is considered as a true positive (TP) if we can observe unexpected visual effects that can affect apps' visual consistencies and functionalities based on the above reproduction steps, otherwise we consider it as a false positive (FP). 

\begin{figure}[t]
\centering
\includegraphics[width=0.45\textwidth]{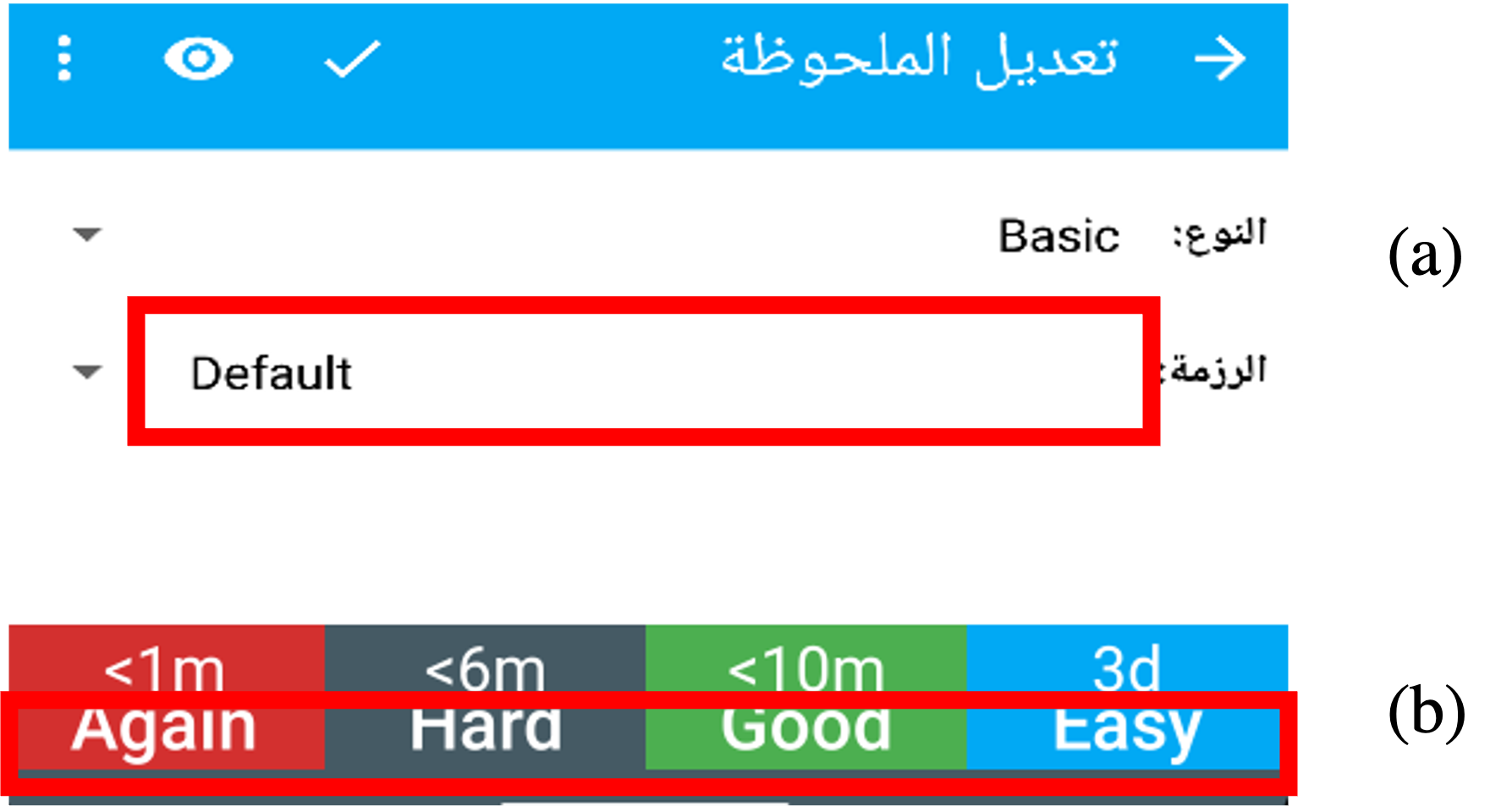}
\caption{Two true positives generated by SUDFinder in AnkiDroid.}
\label{fig:true_positive}
\end{figure}

\subsubsection{Results} As shown in Table~\ref{tab:rq2}, {SUDFinder} detected 129 bugs in 11 subjects, among which 98 are true positives (i.e., the precision is 0.76). This shows that {SUDFinder} can precisely detect SUD bugs in Android apps. On average, SUDFinder found 116.3 TextView and EditText elements per app, and generating text for each TextView or EditText consumed an average of 463.8 tokens (minimum: 298.1; maximum: 824.2). From Table \ref{tab:rq2}, we can see that the SUD bugs detected by SUDFinder are diverse: the apps are affected by different settings with different consequences. We submitted bug reports for each validated TP identified by SUDFinder to the original app developers. Each bug report included detailed steps to reproduce the failure and screenshots to help app developers understand the SUD bugs. We strictly \textit{followed the contributing guidelines and licenses of the app projects when submitting pull requests and bug reports}. So far, 67 bugs have been confirmed and 37 bugs have been merged by the app developers, \textbf{showing the usefulness of SUDFinder}. We also received positive
feedback from developers. For example, one developer of LibreTube comments that ``\textit{Thanks for the detailed report, I can't tell why we previously forced a dark action bar theme there}'';

We conducted three independent repetitions on the 98 true positives identified by SUDFinder. The results show that 95 were consistently detected across all three runs, while the remaining 3 were not detected in every run. These three cases mainly arise because the bug manifestation depends on the generated text spanning multiple lines (e.g., left-aligned formatting and other layout factors), which introduces variability in the trigger conditions. 

We further delve into the app code level of detected true positives and summarize the following two major root causes of SUD bugs. These two root causes account for 69.4\% of the TPs identified.

\begin{itemize}[leftmargin=1ex]
\item We found that 45 TPs occur because app developers assign fixed values to the visual appearance, preventing it from adjusting correctly according to system settings. For example, SUDFinder identified a bug in AnkiDroid where the attribute \texttt{android:textAlignment="inherit"} was specified for the ``Default'' view, causing it to fail in aligning to the right, unlike the ``Basic'' view when switching to Arabic (see Figure \ref{fig:true_positive}(a)).

\item The Android framework provides various attributes that can automatically adjust the visual appearance of UI components based on different system settings. However, we identified 23 cases where app developers used incorrect attributes, resulting in the visual appearance not changing as expected. For instance, SUDFinder found a bug in AnkiDroid where the view becomes partially invisible when the font size is set to the largest, due to the hard-coded attribute \texttt{android:layout\_height="36dp"} (see Figure \ref{fig:true_positive}(b)).

\end{itemize}

\begin{figure}[t]
\centering
\includegraphics[width=0.45\textwidth]{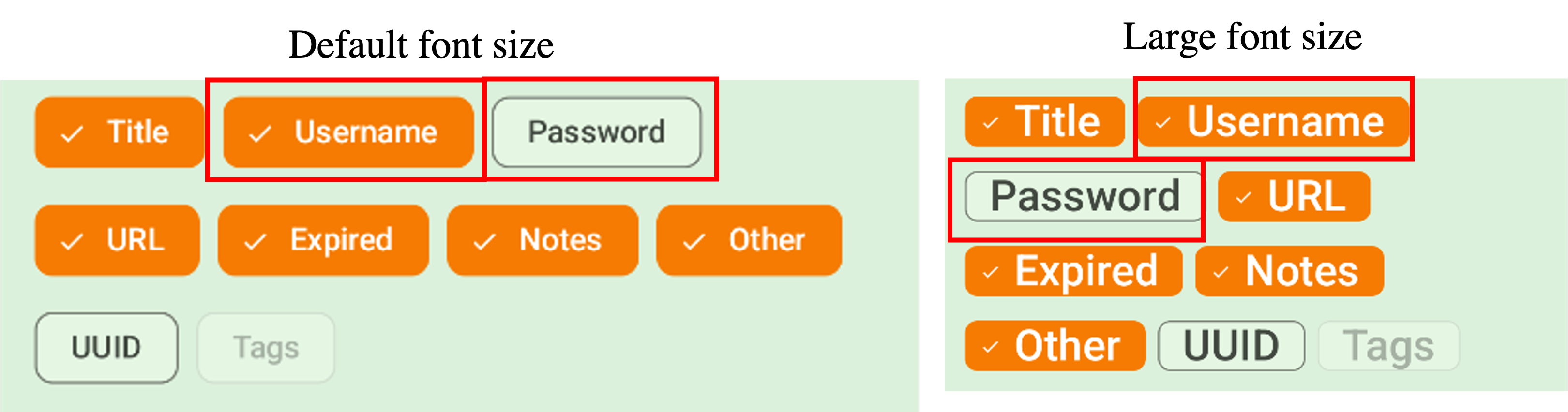}
\caption{A false positive generated by SUDFinder in KeePassDX.}
\label{fig:false_positive}
\end{figure}

\begin{figure}[t]
\centering
\includegraphics[width=0.45\textwidth]{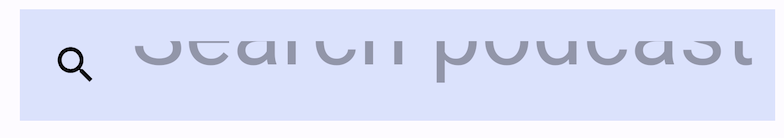}
\caption{A false negative of SUDFinder in AntennaPod.}
\label{fig:false_negative}
\end{figure}

We further investigated 31 FPs to understand the limitations of {SUDFinder}. The main reason is that the breakage of UI style consistency did not affect the app's functionality or disrupt the visual consistency. Figure \ref{fig:false_positive} shows an example of a FP. As discussed in Section~\ref{sec:approach}, SUDFinder checks for the consistency of layout alignment before and after system setting changes. In the example shown in Figure \ref{fig:false_positive}, although ``Username'' and ``Password'' are no longer aligned together after the text scale is doubled, it does not affect the normal use of the app by the user, nor does it impact the aesthetic appearance of the app's UI.

\subsection{RQ6: \revision{Ablation Study}}
\subsubsection{Setup}
To answer RQ6, we conducted the following two experiments to evaluate \revision{(1) the MLLM-based text generation module and (2) the XML-based test app generation strategy (i.e., the state detection capability) of SUDFinder.}

\textbf{MLLM-based Text Generation.} To evaluate the contribution of the MLLM-based text generation module, we compared SUDFinder with SUDFinder$_{nollm}$, a variant where this module is removed while all other functionalities are retained. To further evaluate the robustness and generalizability of the module across different MLLMs, we also tested SUDFinder with two open-source MLLMs (Qwen-VL and GLM-4.1V) in addition to the default GPT-4o. Detailed specifications of the evaluated MLLMs are provided in Table~\ref{tab:mllm}.

\begin{table}[t]
\centering
\caption{Studied Multi-Modal Large Language Models}
\label{tab:mllm}
\begin{tabularx}{\linewidth}{Y Y Y Y}
\toprule
\textbf{Source} & \textbf{Model Name} & \textbf{Company} & \textbf{Size} \\
\midrule
\multirow{2}{*}{Open-source}
    & Qwen-VL    & Alibaba  & 3.3B \\
    & GLM-4.1V   & Zhipu    & 10B \\
\midrule
\multirow{1}{*}{Closed-source}
    & GPT-4o   & OpenAI   & - \\
\bottomrule
\end{tabularx}
\end{table}

\textbf{XML-based App Generation.} To evaluate the contribution of the XML-based test app generation strategy, we compared SUDFinder with SUDFinder$_{r}$, a variant where the XML-based strategy is replaced by random exploration following the practices of Sun et al.~\cite{sun2021understanding, sun2023characterizing} (20 seed tests $\times$ 100 events), while the test oracle is kept unchanged.

\subsubsection{Results}
We report the results of RQ6 in Table~\ref{tab:rq2}.

\textbf{MLLM-based Text Generation.} As for SUDFinder$_{nollm}$, it generates 60 TPs and 23 FPs, with a precision of 0.72. SUDFinder covers all TPs detected by SUDFinder$_{nollm}$; however, SUDFinder$_{nollm}$ fails to detect 38 TPs whose detection relies on MLLM-based text generation. These 38 TPs span T1.1--T1.4 bugs, demonstrating that the MLLM-based text generation module effectively simulates runtime text content and enhances bug detection coverage. Regarding generalizability, SUDFinder detects 96 and 98 TPs when leveraging Qwen-VL and GLM-4.1V, achieving a precision of 0.72 and 0.74, respectively — both close to the GPT-4o baseline. All TPs generated by Qwen-VL and GLM-4.1V are also detected by GPT-4o, showing that SUDFinder's effectiveness generalizes well across different MLLM architectures.

\textbf{XML-based App Generation.} As we can see in Table~\ref{tab:rq2}, SUDFinder$_{r}$ generates 49 TPs and 13 FPs, with a precision of 0.79. Among the 49 TPs detected by SUDFinder$_{r}$, SUDFinder covers 40 (81.6\%), while the remaining 9 are missed mainly because their runtime information is defined in app code rather than XML configuration files. However, the remaining 58 TPs detected by SUDFinder are not covered by SUDFinder$_{r}$, primarily because random exploration fails to reach all UI components defined in the app — for example, exploring certain pages requires triggering specific in-app conditions that random events cannot reproduce. These results indicate that the XML-based test app generation strategy enables SUDFinder to explore more buggy UI components that dynamic random exploration cannot reach.

\subsection{RQ7: Effectiveness of Test Oracle}

To evaluate the effectiveness of the proposed test oracle, we compare SUDFinder with baselines whose tools are publicly available and capable of running on API level 35, the latest Android framework version on December 2024. As illustrated in Section \ref{sec:motivation}, we compare the effectiveness of SUDFinder with the following baselines: SetDroid~\cite{sun2021understanding, sun2023characterizing}, dVermin~\cite{su2022metamorphosis}, ITDroid~\cite{escobar2020empirical}. We followed the original configuration of dVermin to conduct experiments. For SetDroid, except for the display and language settings, we exclude other settings primarily because they rarely cause SUD bugs, as noted in Section \ref{sec:empirical}. We enabled SetDroid to check UI consistencies if a given setting is changed and later properly restored. We did not configure SetDroid to detect SUD bugs that manifest as unexpected differences when settings are not restored, since SetDroid does not account for the expected UI adaptation related to T1 bugs.

\subsubsection{Results}

Table \ref{tab:rq2} results show the effectiveness of the baselines. For SetDroid, it generates 10 TPs in 4 apps, achieving a precision of 0.83. Among these 10 TPs, 5 are app crashes, which cannot be detected by SUDFinder because the crashes are located in the app code, not in the XML files. The remaining 5 are caused by visual features triggered by specific system events. For example, the keyboard of the app droid-ify becomes invisible before and after rotating the screen, leading to some UI inconsistencies between the two devices. These 5 TPs also cannot be found by SUDFinder because the bug detection process does not involve system/user events. However, all the 98 TPs generated by SUDFinder-GPT cannot be detected by SetDroid due to (1) not covering the buggy UI pages, and (2) not modeling the expected changes related to UI styles, as illustrated in Section \ref{sec:empirical}.

For dVermin, it successfully detects 11 TPs related to text containment inconsistencies (T1.4 bugs), achieving a precision of 0.85. Among the 11 TPs found by SUDFinder, dVermin is able to detect 6 of them, while the remaining 5 TPs cannot be detected due to the limitation of random exploration. On the other hand, 5 TPs detected by dVermin cannot be covered by SUDFinder, mainly because SUDFinder fails to accurately restore the boundary of visual content. For example, the bug in Figure \ref{fig:false_negative} cannot be detected by SUDFinder, primarily because the visual content boundary extends beyond the edit text, and the extended part is invisible.

ITDroid successfully detects 17 TPs, all of which are also identified by SUDFinder. Of these, 12 are related to T1.2 bugs caused by layout bound drifts, and 5 are linked to T1.4 bugs. However, as mentioned in Section \ref{sec:motivation}, ITDroid's detection process only considers layout bounds and does not model UI styles related to visual content, such as colors (T1.1 bugs). Consequently, ITDroid fails to detect the other 81 SUD bugs identified by SUDFinder.

The above results show that SUDFinder complements baseline methods. First, SUDFinder focuses on the detection of T1 bugs, which are mainly induced by UI style inconsistencies and require understanding the relations of UI components before and after setting changes. Such relations have not been completed summarized in existing studies~\cite{su2022metamorphosis, sun2021understanding, sun2023characterizing}. Second, directly analyzing the XML configuration files in the app project allows for comprehensive testing of the UIs included in the app, without relying on random testing or manually-built test cases as existing approaches do, thereby helping developers find more previously unknown SUD bugs.

\begin{table}[t]
\centering
\caption{Results of SUDFinder and Baselines}
\label{tab:rq2}
\resizebox{0.49\textwidth}{!}{
\begin{threeparttable}
\begin{tabular}{cccccc}
\toprule
\textbf{Approach} & \textbf{Total} & \textbf{Color} & \textbf{Alignment} & \textbf{Distance} & \textbf{Containment} \\
\midrule
SUDFinder-GPT & 98/129(0.76) & 15/22 & 46/60 & 26/34 & 11/13 \\
SUDFinder-Qwen & 96/134(0.72) & 16/24 & 50/64 & 22/37 & 8/9 \\
SUDFinder-GLM & 98/132(0.74) & 13/18 & 48/62 & 26/40 & 11/12 \\
SUDFinder-GPT$_{r}$ & 49/62(0.79) & 9/14 & 28/35 & 4/4 & 8/9 \\
SUDFinder$_{nollm}$ & 60/83(0.72) & 13/19 & 26/36 & 12/18 & 9/10 \\
SetDroid & 10/12(0.83) & 10/12 & - & - & - \\
dVermin & 11/13(0.85) & - & - & - & 11/13 \\
ITDroid & 17/25(0.68) & - & 12/20 & - & 5/5 \\
\bottomrule
\end{tabular}
\begin{tablenotes}
\footnotesize
\item \textbf{TP/Total(P)}: \textbf{TP} is the number of TPs; \textbf{Total} is the total number of generated bug reports; \textbf{P} is the precision of the specific approach. \textbf{``-''} means the approach is not applicable for detecting the specific bug types.
\end{tablenotes}
\end{threeparttable}
}
\end{table}

\subsection{RQ8: Comparison with MLLM}

\begin{table*}[htbp]
\centering
\caption{Prompt Structure}
\label{tab:prompt_structure}
\small
\begin{tabularx}{\textwidth}{>{\hsize=0.25\hsize}X >{\hsize=1.75\hsize}X}
\toprule
\textbf{Structure} & \textbf{Prompt} \\
\midrule
\textbf{Question}
& Android apps must adapt to various system settings (e.g., language, theme, screen orientation, display size).
Failure to do so results in Setting-Related UI Display bugs.

 1. You will be provided with a screenshot under system setting \textit{(sys\_setting)}.

 2. Analyze the screenshot to determine if any Setting-Related UI Display (SUD) bugs are present based on this context and the rules below\\
\midrule
\textbf{Rules}
& Analyze the screenshot for SUD bugs based on the following types of UI inconsistencies patterns.

- Style-Related (T1):

  1.Color: Check if UI components have poor contrast or unexpected color differences under the current setting.

  2.Alignment: Check if UI components that should be aligned are misaligned.

  3.Distance: Check if the spacing between semantically-related UI components has changed abnormally.

  4.Containment: Check if a UI component is incorrectly clipped, overlapped, or extending beyond the bounds of its parent container.

- Semantic-Related (T2):

  5.Unexpected Visibility: Check if a UI element (e.g., dialog, notification, error message) remains visible when it should have disappeared based on the current system setting or app state.

  6. Data Handling: Check if a UI component fails to properly manage the data displayed in it (e.g., data loss), which can even lead to app crashes.

  7.Translation: Check if text has incorrect linguistic formatting (e.g., wrong word order), inaccurate translation, or remains untranslated after a language setting change.\\
\midrule
\textbf{Input}
& \textit{(Screenshot of the Current Page)} \\
\midrule
\textbf{Output format}
& - Verdict: Clearly state 'SUD Bug Found' or 'No SUD Bug Found'.

 - Defect Location: If a bug exists, precisely describe its location (e.g., 'Text in the bottom button is truncated') and the specific manifestation (e.g., 'Layout breaks after language switch').

- Rationale: Briefly explain the reasoning behind your judgment. \\
\bottomrule
\end{tabularx}
\end{table*}

To address RQ8, we evaluated the effectiveness of MLLMs in detecting SUD bugs in Android apps. Following the methodology described in~\cite{ju2024studyofusingmllm}, we designed the prompt by (1) providing the relevant rules and (2) giving the screenshot of the current page. We asked the MLLM to identify the location of the bug and briefly explain the rationale behind its judgment.

We leveraged the MLLMs listed in Table \ref{tab:mllm} for the evaluation. Specifically, for each MLLM, we conducted random exploration of the subject apps \revision{following the methodology of Sun et al.\cite{sun2021understanding, sun2023characterizing}}, generating a total of 174,000 screenshots ($ \text{3 settings} \times \revision{ \text{20 seed tests} \times \text{100 events}} \times \text{29}$ apps). \revision{The substantial volume of screenshots necessitated extensive manual verification. Therefore, we randomly sampled 5 screenshots for each combination of application and system setting. The selected screenshots were based on unique UI states, ensuring that the pixel difference between screenshots exceeded 10\% to guarantee diversity. Using the above sampling method, we ultimately obtained 435 screenshots ($ \text{29 apps} \times \text{3 settings} \times \text{5}$) per MLLM for manual verification. }

\begin{figure}[t]
\centering
\includegraphics[width=0.35\textwidth]{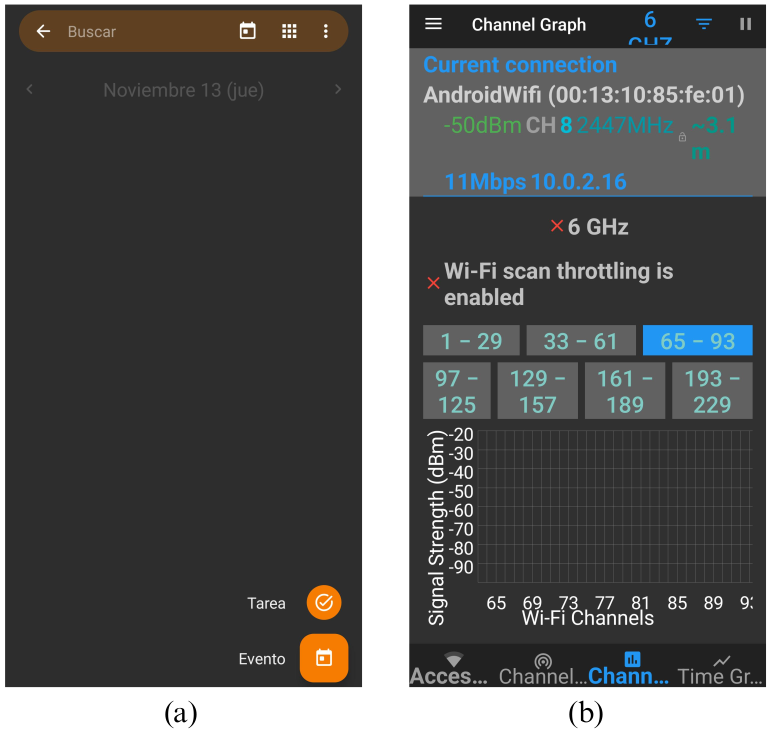}
\caption{Two false positives generated by MLLMs.}
\label{fig:mllm_false_positive}
\end{figure}

The results are shown in Table \ref{tab:mllmresult_new}.
GPT-4o can only achieve a precision of 0.09 by generating 24 TPs and 249 \revision{FPs}, while Qwen-VL and GLM-4.1V scored 0.21 (44 TPs and 164 \revision{FPs}) and 0.12 (38 TPs and 286 \revision{FPs}), respectively.
As mentioned in Section \ref{sec:motivation}, although MLLM (such as GPT-4o) can detect bugs using app context provided in screenshots, its knowledge, derived from large-scale corpora, does not effectively model SUD bugs. \revision{As shown in Figure~\ref{fig:mllm_false_positive}, MLLMs tend to misidentify UI content unrelated to SUD as SUD bugs. In Figure~\ref{fig:mllm_false_positive}(a), no actual SUD bug exists, yet after the app switches to the dark theme, the MLLM misreports the blank region as a defect, producing the output: \textit{``Defect Location: Main content area remains blank after applying dark theme. Rationale: The primary content area fails to display expected UI elements, indicating a failure to adapt to the system's dark theme setting, which is a SUD bug.''} However, the blank region merely reflects an empty application data state (no events or tasks currently exist) that would appear identically under the default theme, revealing that the MLLM has conflated application-level data state with style-level rendering behaviour. In Figure~\ref{fig:mllm_false_positive}(b), although a SUD bug is present, the MLLM additionally misreports an intrinsic overflow-handling behaviour in the bottom navigation bar as a defect, producing the output: \textit{``Defect Location: Text in the bottom navigation bar is truncated (e.g., `Acces...' instead of `Access Points'), with the specific manifestation of layout breaking due to font scale. Rationale: Under font scale 2.0, text extends beyond its container boundaries, violating the `Containment' sub-rule under Style-Related (T1).''} Yet this truncation already exists under the default font scale and is an inherent overflow-handling mechanism of the component. The root cause of both false positives is that the MLLM reasons solely over the post-setting screenshot without any baseline comparison, making it unable to distinguish setting-induced regressions from pre-existing, by-design layout behaviours.}
The above results demonstrate that while MLLM has the potential to detect SUD bugs, the significant number of \revision{FPs} reduces its practical effectiveness in identifying real-world SUD bugs.

In terms of API overhead, GPT-4o consumed an average of 2,434 tokens (minimum: 2,179; maximum: 2,863), while Qwen-VL required an average of 3,153 tokens (minimum: 3,048; maximum: 3,943), and GLM-4.1V used an average of 6,496 tokens (minimum: 6,067; maximum: 7,245).

\begin{table}[t]
	\centering
	\caption{Results of MLLM in Detecting SUD Bugs}
	\label{tab:mllmresult_new}
	\resizebox{0.45\textwidth}{!}{
		\begin{threeparttable}
			\begin{tabular}{cccccc}
				\toprule
				\textbf{Model} & \textbf{Total} & \textbf{Color} & \textbf{Alignment} & \textbf{Distance} & \textbf{Containment} \\
				\midrule
				Qwen-VL  & 44/208(0.21) & 8 & 16 & 0 & 20 \\
				GLM-4.1V & 38/324(0.12) & 9 & 13 & 0 & 16 \\
				GPT-4o   & 24/273(0.09) & 5 & 4  & 0 & 15 \\
				\bottomrule
			\end{tabular}
		\end{threeparttable}
	}
\end{table}

\section{\revision{Discussion}}
\label{sec:discussion}

\subsection{\revision{Contribution to Research}}
\revision{This work contributes to the research on setting-related UI testing in two aspects. First, we extend the empirical dataset of Sun et al.~\cite{sun2021understanding, sun2023characterizing} and provide the first systematic taxonomy of setting-related UI display bugs (T1 style-related and T2 semantic-related, with four T1 sub-categories), together with their triggering settings, code-level locations, and user-facing consequences. Second, we introduce XML-driven test-activity injection as a coverage mechanism for SUD bug detection. To the best of our knowledge, this is the first approach that force-renders each XML layout through an injected activity, enabling detection on UI pages that dynamic event-driven exploration cannot reach.}

\subsection{\revision{Contribution to Practice}}
\revision{The results of this work are directly usable by Android practitioners. First, SUDFinder is released as an open-source tool that has already reported 98 previously unknown SUD bugs to 29 actively maintained apps, of which 67 have been confirmed and 37 have been merged by the original developers. Second, the T1 test-oracle designs (color, alignment, distance, and containment consistency across settings) are decoupled from our implementation and can be adopted by any UI testing framework to extend its setting-related bug detection capability. Third, we release the labeled dataset of 308 SUD bugs, which practitioners can use as regression cases and researchers can use as an evaluation benchmark. Finally, the common triggering patterns summarized in our empirical study (e.g., dark mode, font scaling, landscape mode, and internationalization/localization) highlight concrete review points that app developers can apply during code and PR review to prevent SUD bugs early.}

\subsection{Threats to Validity}
\label{sec:threats}
\textbf{Empirical Dataset Selection. }
The empirical study relies on the state-of-the-art empirical dataset released in January 2021. \revision{To mitigate the potential limitation of dataset recency, we have extended the original dataset with 90 additional SUD bugs collected from actively maintained Android apps up to the submission date of this paper. These additional bugs were collected and validated following the same methodology as the original dataset, ensuring consistency and quality. While the combined dataset may still not cover all the most recent SUD bugs, our experimental results demonstrate that the insights gained from it are effective in detecting previously unidentified SUD bugs in actively maintained apps.}

\textbf{Evaluation Subject Selection.} The findings from our evaluation depend significantly on the representativeness of the chosen subjects. To address this concern, we selected 29 real-world, open-source Android apps that are large-scale, actively maintained, and varied across different categories. We also followed the evaluation settings of baselines when conducting experiments.

\textbf{Subjectivity of Manual Inspections.}
We manually reviewed SUD bugs in the empirical dataset to derive our findings and validated bug reports generated by {SUDFinder} and baseline methods through manual inspection. To reduce errors in manual process, two authors collaborated to reach a consensus, ensuring consistency. Additionally, we sought feedback from the original app developers and have made both empirical and evaluation datasets available on our project website.

\textbf{Non-Determinism of MLLM.}
\revision{A threat to the reproducibility of our experiments stems from the inherent non-determinism of MLLM outputs. We repeated the experiments three times to mitigate this threat.}

\section{Related Work}
\label{sec:relatedwork}

\subsection{UI Testing in Android Apps}
There are already a number of empirical studies aimed at summarizing different types of UI bugs in Android apps~\cite{huang2021characterizing, hu2018tale, liu2020owl, escobar2020empirical, chen2021accessible, sun2021understanding, sun2023characterizing, huang2018understanding, wei2018understanding}. Note that the empirical study conducted by Sun et al.~\cite{sun2021understanding, sun2023characterizing} illustrates the potential UI display bugs in Android apps. However, they lack a detailed categorization of the root causes behind these UI display bugs. This omission hinders automated tools—which are built on empirical findings—from effectively detecting bugs within system settings. To mitigate this research gap, we conduct an empirical study to understand the common root causes of UI display bugs and propose an automatic approach, SUDFinder, to help detect such bugs.

A number of automated UI testing techniques have been proposed~\cite{mirzaei2015sig, gu2019practical,pan2020reinforcement,dong2020time,su2016fsmdroid}. Such UI testing techniques can be categorized into two types. The first one is the model-based approaches~\cite{azim2013targeted, su2017guided, lv2022fastbot2, pan2020reinforcement, baek2016automated, degott2019learning} with the aim of exploring
diverse UI states of an Android app. Another type of approaches take the specific program properties (e.g., branches, API invocations) to generate UI test cases~\cite{mao2016sapienz,wang2020combodroid, su2017guided}.
In view of the rapid development of UI testing approaches in Android apps, there has been related work studying the practical effectiveness and challenges of these automated tools in industry~\cite{wang2018empirical, su2021benchmarking, choudhary2015automated}. However, these testing
techniques are limited to crash defects due to lack of strong
test oracles. On the other hand, our proposed approach, SUDFinder, targets SUD bugs that affect apps' UI and rarely induce app crashes.  

In addition, research work has been proposed to test bugs related to UI display in Android apps~\cite{huang2021characterizing, liu2020owl, su2022metamorphosis, hu2018tale, escobar2020empirical}. Specifically, Liu et al.\cite{liu2020owl} proposed OwlEye, an automated approach to detect UI display bugs based on a large dataset of real-world UI display bugs. Huang et al.\cite{huang2021characterizing, huang2023conffix} proposed ConfDroid and ConfFix, aiming to detect and fix compatibility issues induced by apps' XML configuration files, which are widely used to design apps' UI. Su et al.\cite{su2022metamorphosis} proposed dVermin, an automated approach to detect UI display issues under the system setting of text scaling. Hu et al.\cite{hu2018tale} proposed wDroid, an automated approach to detect UI display bugs related to WebViews in Android apps. Fazzini et al.~\cite{fazzini2017automated} proposed DiffDroid, which conducts differential testing to detect UI incompatibilities. Different from the above approaches, we target UI display bugs related to different types of system settings. Our study involves new types of bugs that have not been covered by existing work, and we propose a set of UI component exploration strategies to enhance the bug detection capability of automated approaches. \revision{In particular, these existing approaches~\cite{sun2021understanding, sun2023characterizing, su2022metamorphosis, escobar2020empirical, liu2020owl, hu2018tale} all rely on dynamic event-driven exploration and therefore cannot reach UI pages whose rendering is gated by runtime state (e.g., authentication or server-fetched content); to the best of our knowledge, SUDFinder is the first to detect SUD bugs via XML-driven test-activity injection, which force-renders each layout to cover these otherwise-unreachable pages.}

\subsection{Setting-Related Bugs in Android Apps} 

There are also existing research efforts towards setting-related bugs in Android app. Some existing studies have proposed the automatic detection and repair approaches for specific types of setting defects (e.g., permissions~\cite{sadeghi2017patdroid}, app-specific preference~\cite{lu2019preference}, data loss~\cite{guo2022detecting, zhou2023ddldroid, riganelli2020data}, font scaling~\cite{su2022metamorphosis}, mobile phone accessibilities~\cite{eler2018automated, salehnamadi2021latte}). In addition to these studies, Sun et al.~\cite{sun2021understanding, sun2023characterizing} introduced SetDroid and SetChecker as state-of-the-art approaches for automatically detecting and repairing various setting defects in Android apps. Specifically, SetChecker conducts static analysis and integrates two major fault patterns to detect setting defects at the code level. SetDroid employs a metamorphic relation to conduct dynamic analysis. This relation asserts that altering the system settings and promptly restoring them should have no impact on the subsequent execution of the apps. However, SUDFinder addresses the limitations of monitoring how app components adapt their runtime behavior in response to setting changes. As a result, it can uncover setting-related bugs that SetDroid and SetChecker are unable to detect. As demonstrated in the study by Sun et al.~\cite{sun2023characterizing}, different types of setting-related bugs require different detection strategies. We believe that SUDFinder can serve as a complement to the existing approaches.

\section{Conclusion}

In this paper, we conducted an empirical study aimed at understanding the common root causes and patterns of setting-related UI bugs (SUD bugs) in Android apps. Based on our findings, we proposed SUDFinder, a tool that encodes common bug patterns to automatically detect SUD bugs in Android apps. Our results demonstrate that SUDFinder outperforms existing baselines in accurately identifying real and previously unknown SUD bugs. In the future, we plan to explore the automatic repair of such SUD bugs in Android apps.

\section{Data Availability}
The implementation of SUDFinder, along with the empirical and evaluation datasets, is available at \url{https://github.com/getgo-nobugs/SUDFinder/}.

\section*{Acknowledgment}
This work was supported by the National Natural Science Foundation of China (Grant Nos. 62272400, 62402405), Fujian Provincial Natural Science Foundation (Grant Nos. 2025J010002, 2026J001003), Xiamen Natural Science Foundation (Grant No. 3502Z202471016), and the Fundamental Research Funds for the Central Universities (Grant Nos. 20720240087, 20720250029).

\bibliographystyle{IEEEtran}
\bibliography{reference}

\raggedbottom

\begin{IEEEbiography}[{\includegraphics[width=1in,height=1.25in,clip,keepaspectratio]{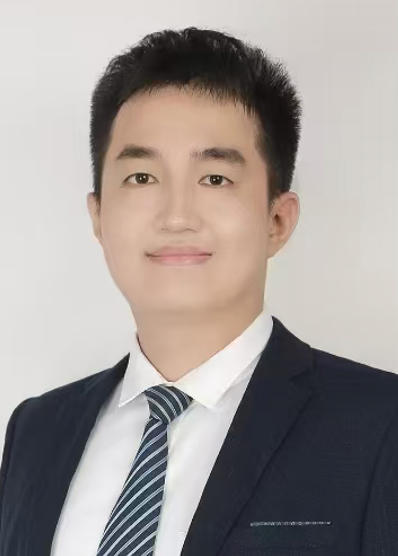}}]{Huaxun Huang}
received the PhD degree from HKUST in 2023. He is currently an assistant professor at  School of Informatics, Xiamen University. His research interests include software engineering, program analysis, and mobile application testing. His research work has been regularly published in top conferences and journals in the software engineering community, including FSE, ASE, ISSTA, TSE, TOSEM, and OOPSLA. More information about him can be found at: https://huaxunhuang.github.io/
\end{IEEEbiography}
\begin{IEEEbiography}[{\includegraphics[width=1in,height=1.25in,clip,keepaspectratio]{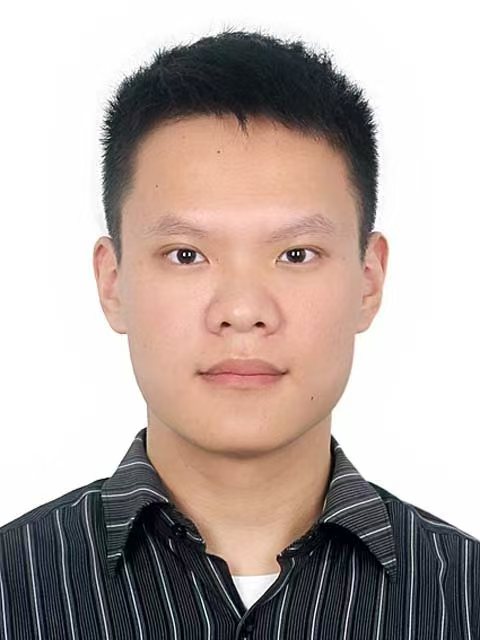}}]{Wu Liu} received the Master's degree from the School of Informatics, Xiamen University, in 2026, and the Bachelor's degree in engineering from Xiamen University in 2023. His research interests include software engineering, with a focus on mobile app testing.
\end{IEEEbiography}
\begin{IEEEbiography}[{\includegraphics[width=1in,height=1.25in,clip,keepaspectratio]{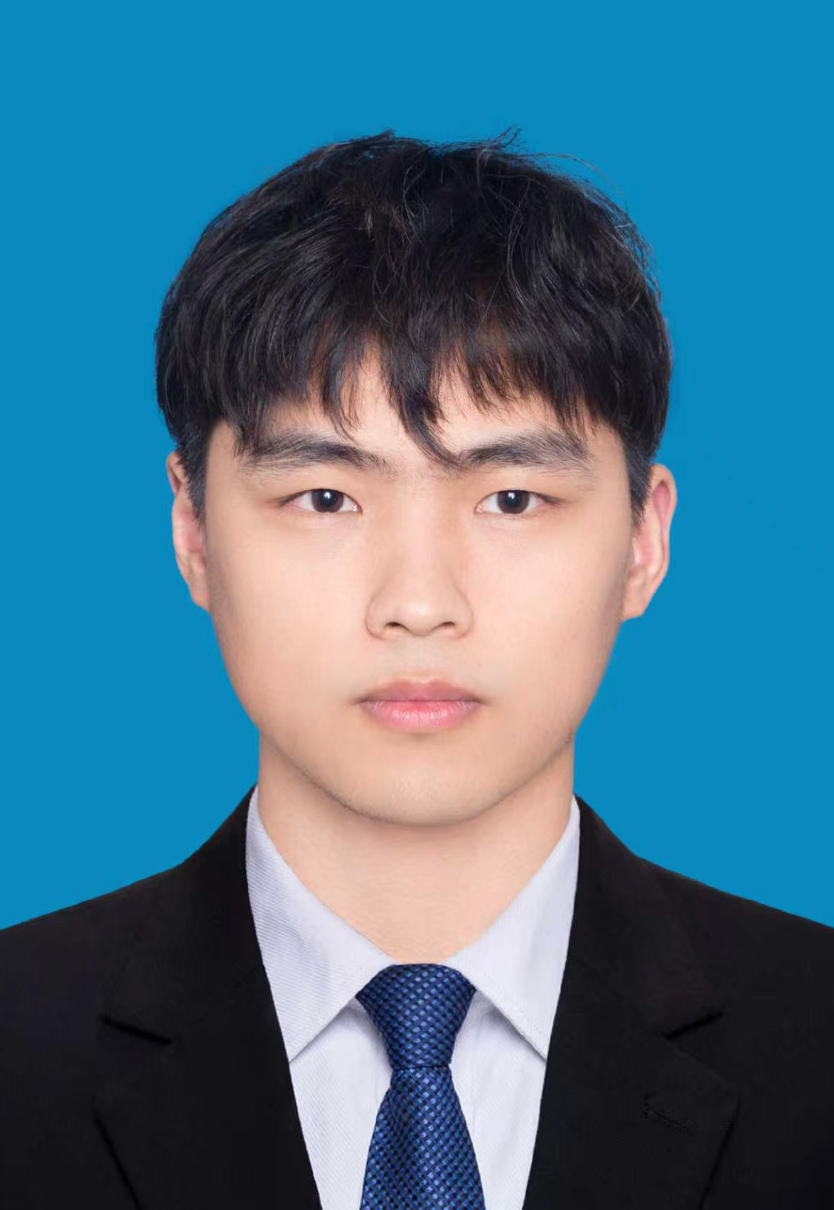}}]{Jiahao Gu} is currently working toward the Master's degree at the School of Informatics, Xiamen University. He received the Bachelor's degree in engineering from Xiamen University in 2024. His research interests include the intersection of artificial intelligence and software engineering. He has published several papers in top-tier conferences and journals, including ASE and TSE.
\end{IEEEbiography}
\begin{IEEEbiography}[{\includegraphics[width=1in,height=1.25in,clip,keepaspectratio]{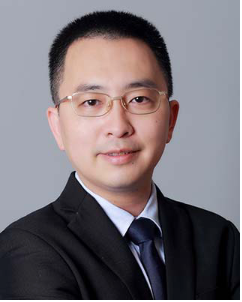}}]{Rongxin Wu} received the PhD degree from HKUST, in 2017. He is currently a full professor at the Department of Computer Science and Technology, School of Informatics, Xiamen University. His research interests include program analysis, software security, and mining software repository. His research work has been regularly published in top conferences and journals in the research communities of program languages and software engineering, including POPL, PLDI, OSDI, ATC, ICSE, FSE, ISSTA, ASE, and TSE and so on. He has served as the associate editor in IEEE Transactions on Software Engineering and a program committee member in several international conferences (ICSE'27, FSE'27, FSE'25, ISSTA'25, SANER'25, FSE'24, ISSTA'24, ASE'23, SANER'23, ASE‘21 and so on). He is a two-time recipient of the ACM SIGSOFT Distinguished Paper Award. More information about him can be found at: https://wurongxin1987.github.io/wurongxin.xmu.edu.cn/
\end{IEEEbiography}

\end{document}